**Advancing Highway Work Zone Safety: A Comprehensive Review of Sensor Technologies for Intrusion and Proximity Hazards**

**Ayenew Yihune Demeke**
Department of Civil, Construction and Environmental Engineering
North Dakota State University, Fargo, North Dakota, 58102
Email: ayenew.demeke@ndsu.edu

**Moein Younesi Heravi**
Department of Civil, Construction and Environmental Engineering
North Dakota State University, Fargo, North Dakota, 58102
Email: moein.younesiheravi@ndsu.edu

**Israt Sharmin Dola**
Department of Mechanical Engineering
North Dakota State University, Fargo, North Dakota, 58102
Email: israt.dola@ndsu.edu

**Youjin Jang (Corresponding Author)**
Department of Civil, Construction and Environmental Engineering
North Dakota State University, Fargo, North Dakota, 58102
Email: y.jang@ndsu.edu

**Chau Le**
Department of Civil, Construction and Environmental Engineering
North Dakota State University, Fargo, North Dakota, 58102
Email: chau.le@ndsu.edu

**Inbae Jeong**
Department of Mechanical Engineering
North Dakota State University, Fargo, North Dakota, 58102
Email: inbae.jeong@ndsu.edu

**Zhibin Lin**
Department of Civil, Construction and Environmental Engineering
North Dakota State University, Fargo, North Dakota, 58102
Email: zhibin.lin@ndsu.edu

**Danling Wang**
Department of Electrical and Computer Engineering
North Dakota State University, Fargo, North Dakota, 58102
Email: danling.wang@ndsu.edu

**ABSTRACT**

Highway work zones are critical areas where accidents frequently occur, often due to the proximity of workers to heavy machinery and ongoing traffic. With technological advancements in sensor technologies and the Internet of Things, promising solutions are emerging to address these safety concerns. This paper provides a systematic review of existing studies on the application of sensor technologies in enhancing highway work zone safety, particularly in preventing intrusion and proximity hazards. Following the Preferred Reporting Items for Systematic Review and Meta-Analyses (PRISMA) protocol, the review examines a broad spectrum of publications on various sensor technologies, including GPS, radar, laser, infrared, RFID, Bluetooth, ultrasonic, and infrared sensors, detailing their application in reducing intrusion and proximity incidents. The review also assesses these technologies in terms of their accuracy, range, power consumption, cost, and user-friendliness, with a specific emphasis on their suitability for highway work zones. The findings highlight the potential of sensor technologies to significantly enhance work zone safety. As there are a wide range of sensor technologies to choose from, the review also revealed that selection of sensors for a particular application needs careful consideration of different factors. Finally, while sensor technologies offer promising solutions for enhancing highway work zone safety, their effective implementation requires comprehensive consideration of various factors beyond technological capabilities, including developing integrated, cost-effective, user-friendly, and secure systems, and creating regulatory frameworks to support the rapid development of these technologies.

## INTRODUCTION

Highway work zones play a critical role in the maintenance and expansion of transportation infrastructures, yet they simultaneously present substantial safety hazards. These zones are unique intersections of workers, construction equipment, and ongoing traffic, creating a dynamic and hazardous environment. Records indicate that between 1.5% and 3% of all yearly workplace fatalities in the United States are attributed to road construction workers *(1)*. Data from the National Institute for Occupational Safety and Health (NIOSH) reveals that from 1982 to 2020, a total of 29,493 individuals, averaging around 776 per year, were fatally injured in work zone incidents across the United States *(2)*. While this number includes drivers and passengers, a significant proportion of these fatalities were construction workers within work zones. According to the same source, out of the fatalities between 2003 to 2020, 2,222 were construction workers who lost their lives at work zones *(2)*. The majority of these accidents typically occur when workers are struck by vehicles. Among the transport-related accidents at work zones from 2011 to 2020, 63% involved workers being struck by vehicles, either intruding from outside or operating within the work zone, resulting in 577 deaths *(2)*. These statistics underscore the critical need for the implementation of effective safety measures in highway work zones to mitigate these risks.

Recognizing the paramount importance of safety in these zones, various conventional measures have been implemented to mitigate potential accidents. Such measures include the deployment of traffic cones and barrels, the use of flaggers, and the enforcement of reduced speed limits around and within work zones *(3)*. Additionally, work zones are typically designed to include safety buffers on both the longitudinal and lateral sides, separating the active work area from moving traffic *(3)*. Despite these ongoing efforts, the rate of accidents and fatalities remained high and showed an increasing trend, as such traditional measures often fall short in addressing the unique nature of these environments. According to the National Safety Council, 954 people were killed and 42,151 people were injured in work zone crashes in 2021, which represents a 63% increase in work zone deaths since 2010 *(4)*. Moreover, data from the Bureau of Labor Statistics highlights a growing concern that work zone fatalities rose from 756 in 2018 to 954 in 2021 *(5)*. With the rapid advancement of sensor technologies, there is a significant potential to mitigate accidents in highway work zones by proactively detecting safety risks and promptly alerting drivers and workers *(3)*. A number of studies have been conducted in this regard, leading to the development of diverse sensor-based systems for vehicle intrusion and proximity detection, each featuring a unique architecture. However, a comprehensive understanding is still lacking regarding the wide array of sensors available for such applications, their suitability for various scenarios, and their comparative advantages and disadvantages. Additionally, there is a gap in knowledge about how these sensor technologies can be systematically and optimally deployed for maximum safety benefit *(3)*. The primary objective of this review paper is to conduct an extensive exploration and synthesis of contemporary literature on the application of sensor technologies for reducing intrusion and proximity incidents in highway work zones. By examining the mechanisms and applications of various sensor technologies, this review aims to provide a comprehensive understanding of their capabilities, advantages, and limitations, as well as investigate their performance characteristics and the additional considerations necessary for their effective adoption.

While there have been efforts to compile review papers in this regard, most have concentrated on general construction sites, often overlooking the specific challenges inherent to highway work zones. Unlike other construction sites, highway work zones are defined by a set of distinct conditions. Firstly, they are characterized by the presence of live traffic, necessitating specific safety measures against vehicle intrusion *(6,7)*. Secondly, the typical noise of construction sites is further amplified by the sounds from moving vehicles, which requires special consideration when selecting sensor technologies *(6–8)*. Thirdly, unlike stationary construction sites, highway work zones are often temporary and mobile, with work shifting periodically along road segments *(7)*. Moreover, many highway work zones are situated in areas where a direct power supply is not readily accessible, making the choice of power source and consumption a significant factor in selecting appropriate sensors. Lastly, the operational areas within these zones are frequently confined due to the need to maintain traffic flow, presenting additional challenges *(7)*. This paper presents a detailed review of sensor technologies in highway work zones, specifically addressing these unique challenges and characteristics. It primarily concentrates on examining sensor technologies identified in current research as appropriate for use in highway work zone environments. Emphasis is placed on location and motion sensors, which play a crucial role in detecting and alerting against proximity hazards and vehicle intrusions.

This review follows the Preferred Reporting Items for Systematic Review and Meta-Analyses (PRISMA) protocol. The methodology involves a detailed literature search, selection based on pre-defined criteria, and a thorough evaluation of the selected studies. This process ensures the inclusion of relevant, high-quality studies, enabling a comprehensive overview of the current state of sensor technology in highway work zone safety. The results of this review could be helpful in various ways. Understanding the diverse array of sensor technologies employed in highway work zones and assessing their effectiveness provides valuable insights for informed decision-making regarding future sensor applications. The review also offers a detailed exploration of the distinct roles these sensors play in work zone environments and evaluates their relative suitability. By doing so, stakeholders can make better choices in aligning the right sensor technology with the intended purpose. Moreover, this review provides valuable recommendations on the successful adoption of these technologies and important factors to consider for effectively harnessing their benefits.

The paper is structured into several sections to provide a thorough understanding of the topic. The methodology adopted in this study is provided after this introduction, which details the systematic review process. This is followed by a section exploring the application of sensor technologies in highway work zones, specifically for localization and detection of intrusion and proximity incidents. The subsequent sections discuss the performance characteristics of these sensor technologies, the challenges in adopting them, and potential future research directions. The paper concludes with a summary of the main findings.

## REVIEW SCOPE AND METHOD

This research aims to contribute to the identification of potential areas of interest for both practitioners and researchers. Systematic literature reviews offer a structured approach to identifying pertinent studies, summarizing their outcomes, critically evaluating their methodologies, and offering insights and recommendations for future research endeavors *(9)*. The systematic review methodology has gained substantial traction across various disciplines, encompassing fields such as medicine and healthcare *(10)*, education and learning *(11)*, entrepreneurship and management *(12,13)*, computer science and software engineering *(14,15)*, and building and construction *(16,17)*. This review approach is valued for its explicit and reproducible nature *(18)*. To facilitate the identification of sensor technology applications, this systematic review will provide a comprehensive overview of sensors, their definitions, characteristics, prevailing trends in the literature, and existing gaps in the study of sensor-based systems in construction work zones, particularly within the highway sector.

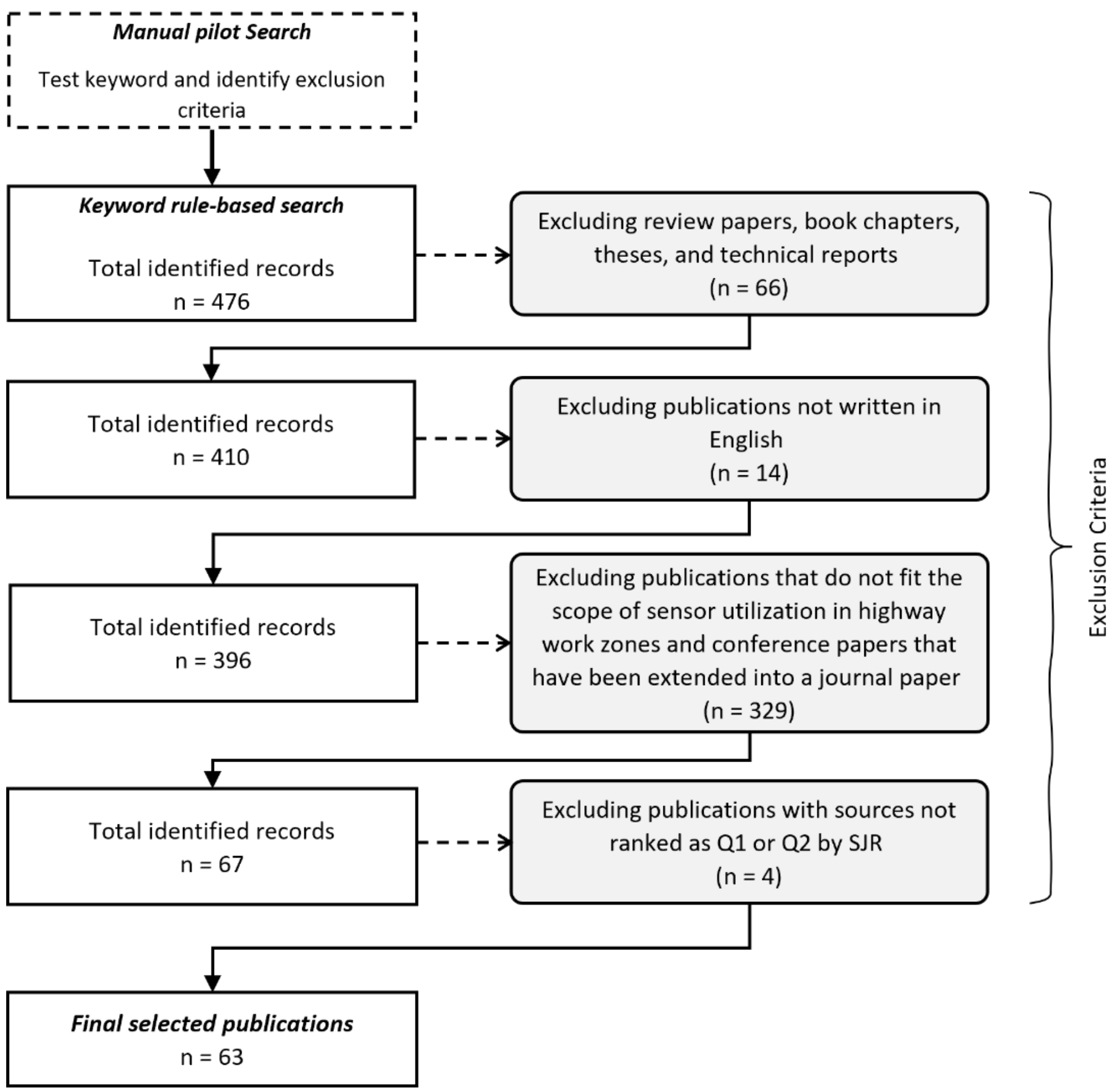

**Figure 1.** Illustration of the systematic publication selection process based on PRISMA.

In order to conduct this comprehensive review, a systematic literature search was performed following the guidelines of the Preferred Reporting Items for Systematic Review and Meta-Analyses (PRISMA) protocol. The utilization of PRISMA not only ensures the production of evidence-based results but also enhances the transparency of the literature selection process *(19)*. One of the key reasons for selecting the PRISMA protocol over other systematic approaches is its methodological clarity and ease of comprehension *(20)*. Furthermore, the adoption of a similar approach has been observed in previous review papers within the field *(21–23)*. A flowchart illustrating the PRISMA phases of this study can be found in Figure 1, and a detailed explanation of the process is provided below.

**Selection Criteria**

In this review, the literature included publications that focus on the deployment of systems incorporating one or multiple types of sensors, with a specific emphasis on addressing the persistent issue of accidents within highway construction work zones. To systematically select the most pertinent publications aligning with the defined subject and scope, a review protocol was established. This protocol comprises inclusion and exclusion criteria, which are outlined in this section.

The following inclusion criteria were applied:

1. Only studies published between 2010 and 2023 were considered. This timeframe was selected to ensure the inclusion of recent and relevant publications, offering insights into the contemporary state of the research topic.
2. Only peer-reviewed publications were included, as they undergo rigorous scrutiny, making it possible to thoroughly assess their research methodologies and objectives.
3. While the primary focus of the review is on highway work zones, the first round of publication selection was not restricted to this area and also encompassed diverse domains, such as computer science and mechanical engineering.
4. Only literature published in the English language was included.

The following exclusion criteria were applied:

1. Publications that were not included in conference proceedings or peer-reviewed journals were excluded.
2. Research with the primary objective of analyzing environmental and behavioral work zone risk factors or developing predictive models related to severity, traffic delay, or similar parameters was not included. Additionally, studies involving experiments conducted using simulators or Virtual Reality (VR) were excluded, as the main focus of this review paper is on real-world applications of sensing devices and systems.
3. Review papers, dissertations, and book chapters were excluded from the selection process.
4. Technical reports, including those from Departments of Transportations (DOTs), were not considered for inclusion. It is important to note that while DOTs often conduct research studies and generate project reports in this field, journal or conference papers are usually published based on the findings from these technical reports.
5. Conference papers containing identical content to journal papers were excluded from the analysis, due to the comprehensive coverage and provision of more detailed information often found in journal publications.

**Systematic Review Strategy**

The reference collection strategy was designed based on a systematic search of academic journals, utilizing specific keywords and adhering to the defined time span as outlined in the selection criteria. To ensure comprehensive coverage, similar to the approach adopted by Kim and Kim *(24)*, a combination of automated and manual search methods was employed, utilizing two prominent online academic document search engines. For the automated search, Scopus (https://www.scopus.com/) was utilized due to its extensive coverage of various publication databases and its robust system for defining a complete set of search rules and filters. Scopus incorporates a powerful search engine, ensuring that the data returned aligns with the specified key terms and research scopes *(24)*. Additionally, the Google Scholar search engine (https://scholar.google.com/) was employed for supplementary manual searches to guarantee the inclusion of all relevant literature. To select the most appropriate keywords, a pilot testing phase was carried out in Google Scholar with the initial keywords mainly express work zone smart hazard prevention systems (e.g., intrusion detection, collision avoidance, alert system), and working venue (e.g., construction work zone). During this phase, keywords were iteratively tested and replaced with alternatives from a pool of words related to highway construction sites sensing system, to encompass as much relevant literature as possible. Subsequently, for the automated rule-based search in the Scopus database, the following set of keywords and operators were applied to search within the title, abstract, and keywords of the papers:

(highway OR roadway OR road) AND ("work zone" OR work-zone OR jobsite) AND (safety OR alert OR hazard OR warning OR collision OR intrusion OR crash OR proximity OR localization)

It is important to highlight that while the focus of this review paper revolves around the application of sensing technologies, the initial automated pilot search revealed that incorporating keywords directly related to that (e.g., sensing, wearable, technology) might result in the omission of some relevant papers. This omission occurs when these specific terms are not present in the title, abstract, or keywords of certain papers. Based on the chosen key terms, the initial search yielded a total of 476 studies published between January 2010 and October 2023. Subsequently, 410 peer-reviewed conference and journal articles were retained, with book chapters, reviews, and other types of publications being excluded. Additionally, 14 papers written in languages other than English were also excluded from the selection.

The remaining publications underwent a screening process based on their titles and abstracts to assess their relevance to the scope of this study. Consistent with the second exclusion criteria mentioned earlier, papers unrelated to the development, application, or testing of a sensor-based system, specifically for functions such as intrusion and near-miss detection, or the localization of workers and equipment for other purposes, were removed from the reference pool. After that, also one round of full-text screening was applied on the remaining publications, since more deep evaluation of the scope of some papers was required. The screening of titles and abstracts, and the second round full-text screening, resulted in a filtering out of 329 publications.

The relevant studies that met the specified inclusion and exclusion criteria underwent a further assessment to gauge their research paper quality and their impact on the academic community. In this study, a journal metric was used to assess the quality of the selected papers, a method that has been demonstrated to be beneficial for enhancing the quality of literature reviews *(25–27)*. Similar to Ayodele et al. *(28)*, this study employed the SCImago Journal Rank (SJR) indicator as part of the quality assessment process to minimize potential discrepancies in study ratings or the inclusion of studies susceptible to bias. The SJR, which is sourced from Scopus, offers several advantages compared to other journal citation indicators. It encompasses a wider array of journals, provides journal rankings for individual years within a specified time frame, conducts more precise analyses based on the year of article publication, encompasses a broader period for citation inclusion, spans a wider range of countries and languages, restricts self-citations, and weights citations according to journal significance and the quality of the citing journal *(29)*. Utilizing a 3-year citation window, SJR classifies journals into four quartiles. In this study, only papers published in journals within the first two quartiles (Q1 and Q2), which represent the top 50% of journals in their respective fields *(30)*, were retained. Through this eligibility process, four papers were excluded from the review.

Finally, a total of 63 publications were selected for inclusion in this review, focusing on the role of sensing technologies in enhancing highway work zone safety. To facilitate content analysis, a Microsoft Excel spreadsheet was utilized to systematically extract essential information from the literature. This information encompassed the publication title, year of publication, type of journal, the country of research origin, and content-related information such as utilized sensor types and their applications. Subsequent to the completion of the data extraction process, a comprehensive content analysis was carried out. The results of this analysis are presented in the following section of this review.

**Statistics of Publications**

This section presents different statistical information about the reviewed publications. Figure 2 illustrates the number of publications per year. The publication trends from 2010 to 2023 indicate a generally increasing interest in the field, despite some year-to-year fluctuations. The publications for this review were sourced from 37 different journals and conference proceedings, with the most notable sources contributing two or more papers, summarized in Table 1. The listed sources contributed more than 60% of the total collection, while Transportation Research Record alone contributed around 16%.

**Table 1.** Top sources of reviewed publications

| Source | Number of papers |
|---|---|
| Transportation Research Record | 10 |
| Automation in Construction | 4 |
| Practice Periodical on Structural Design and Construction | 3 |
| Journal of Construction Engineering and Management | 3 |
| IEEE Conference on Intelligent Transportation Systems Proceedings ITSC | 3 |
| Accident Analysis and Prevention | 3 |
| Transportation Research Part C Emerging Technologies | 2 |
| ISARC 2013 30th International Symposium on Automation and Robotics in Construction and Mining Held in Conjunction with the 23rd World Mining Congress | 2 |
| Journal of Safety Research | 2 |
| Journal of Information Technology in Construction | 2 |
| International Journal of Transportation Science and Technology | 2 |
| Construction Research Congress 2022 Computer Applications Automation and Data Analytics Selected Papers from Construction Research Congress 2022 | 2 |

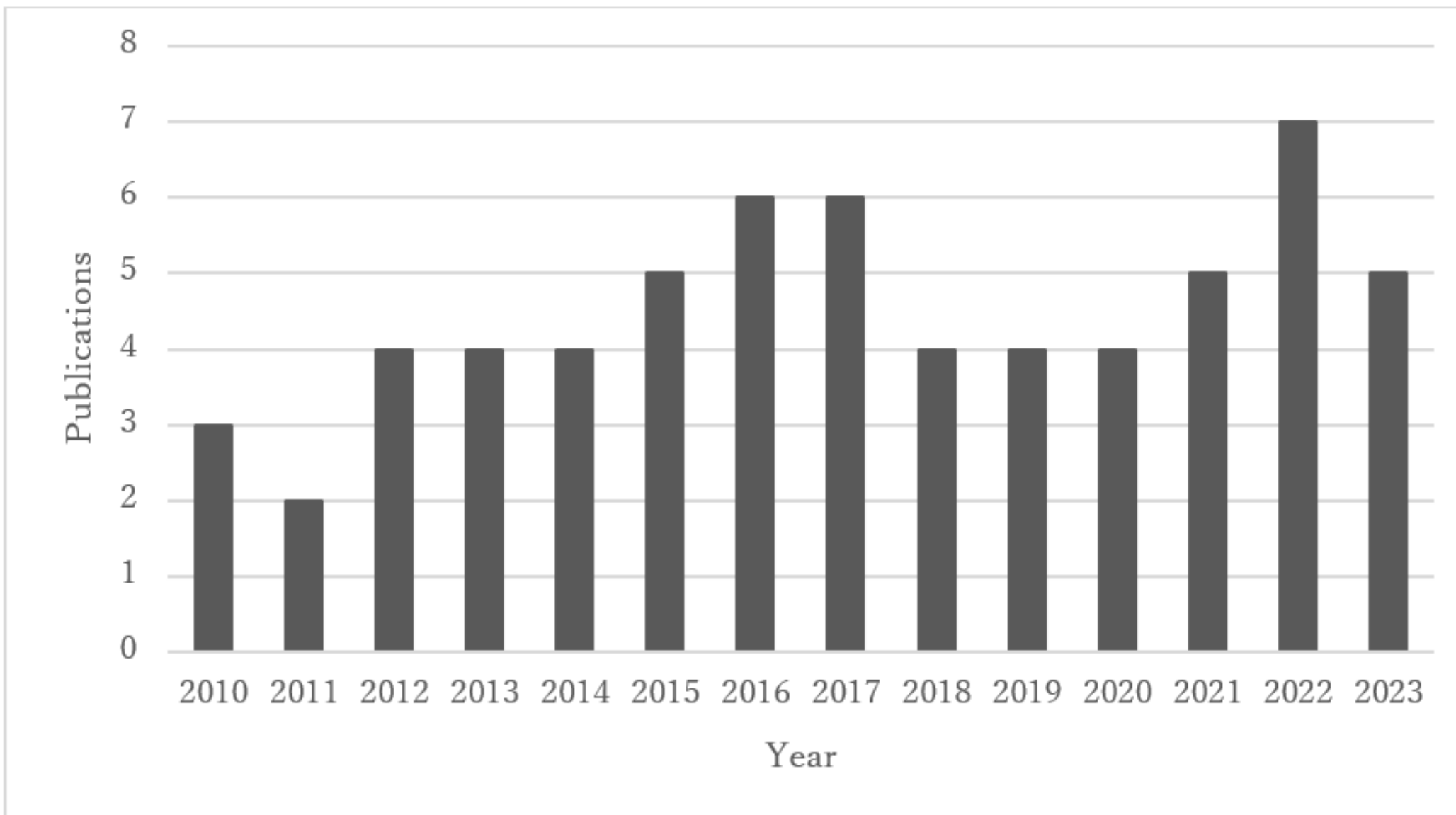

**Figure 2.** Number of publications per year

Additionally, a keyword analysis was conducted using VOSviewer to examine the frequency and network of keywords. The reference list, imported in CSV format from Scopus, was input into the software for this purpose. A co-occurrence analysis of all author keywords was then performed, with a minimum threshold set at one. This process identified 161 keywords, with 106 forming connections. These keywords were visually mapped, as shown in Figure 3. The graph distinctly segregates keywords, with those on the left primarily related to intrusion technologies, and the right highlighting proximity-related keywords and technologies. Notably, the graph indicates “work zone” and “safety” as the two dominant keywords, reflecting the core themes of this review paper.

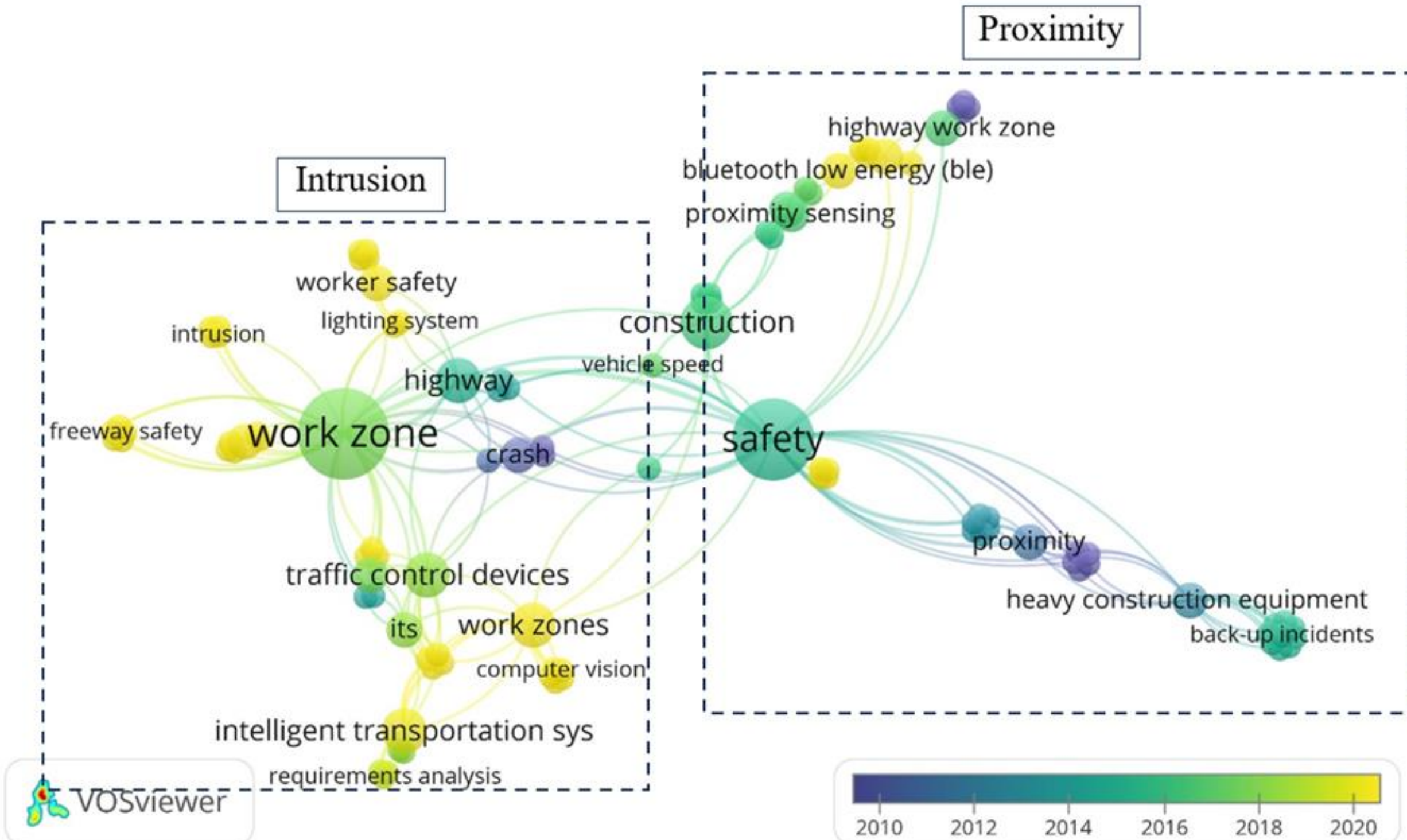

**Figure 3.** Keyword co-occurrence map: VOSviewer version 1.6.20

## APPLICATIONS OF SENSOR TECHNOLOGIES IN HIGHWAY WORK ZONE SAFETY SYSTEMS

The applications of sensor technologies in highway work zones are multifaceted, addressing various critical safety aspects. This section of the manuscript is structured into the application of sensor technologies in three distinct but interconnected areas, each emphasizing a different but crucial safety aspect: localization, vehicle intrusion, and worker-equipment proximity. The interplay of these elements is central to forming a comprehensive safety system. These sections collectively offer a comprehensive overview of the diverse sensor technologies employed and their respective roles in mitigating key safety challenges.

### Localization: Basis for Safety in Work Zones

Localization technology is essential in the array of safety measures for highway work zones, providing the foundational layer upon which systems for intrusion detection and proximity awareness are usually developed *(31–33)*. The primary function of localization is crucial yet straightforward: it accurately tracks and monitors the real-time positions of workers, equipment, and vehicles. This critical data is the cornerstone for implementing effective safety measures, enabling the prompt identification of potential hazards and facilitating swift action in response to emerging risks. The following paragraphs offer a detailed analysis of various sensors utilized for this application, while a summary of key publications in the area is presented in Table 2.

**Table 2.** Summary of key publications on the application of sensor technologies for localization

| Publication | Sensor technologies used | Proposed system | Application |
|---|---|---|---|
| Ibrahim et al. *(34)* | GPS, accelerometer | Low-cost remote sensing hardware | Enables automatic data collection for a near real-time updates on construction equipment activity. |
| Vorhes et al. *(35)* | LiDAR, GPS, accelerometer, gyroscope | A system of low-cost connected work zone devices | Utilizes GPS, accelerometer and gyroscope to track location in a network of connected work zone devices. |
| Nnaji et al. *(36)* | GPS | Wearable Light System | Employs GPS trackers to track worker movements during night-time paving operations. |
| Wang et al. *(37)* | GPS, IMU | A localization and unsafe proximity detection system | Incorporates state tracking and safety rules modules, utilizing GPS and IMU for accurate localization of workers and equipment. |
| Dayal et al. *(32)* | GPS, IMU | A wireless sensor network system | Leverages GPS and IMU with a high-accuracy localization algorithm for precise location determination. |

**Global Positioning System (GPS):** GPS is among the most widely used technologies for outdoor location tracking services *(34,38)*. GPSs use a triangulation method to obtain the position (x, y, z) of a receiver. The position is calculated by measuring the distance from a set of satellites to the GPS receiver, the duration of travel of the GPS signal from satellite to receiver, and the speed of light *(39)*. The main application of GPS in work zone safety systems includes continuously tracking the location of equipment and human objects for various purposes.

In the case of equipment localization and tracking, one notable exploration involves the incorporation of GPS in a low-cost remote sensing hardware system designed to support automated site data acquisition *(34)*. This hardware, driven by an open-source microcontroller (Arduino Uno microcontroller), utilizes GPS for accurate location tracking in outdoor environments. The experiments showcase the versatility of GPS in facilitating near real-time progress reporting of construction operations. The system's capability to detect aggressive driver behavior through a 3D accelerometer, coupled with GPS tracking, exemplifies its adaptability for both equipment and personnel monitoring. In excavator equipment, GPS is employed in conjunction with kinematics to simulate real-world motion within a 3D virtual world *(40)*. In this case, the GPS receiver provides equipment position on a job site, facilitating accurate monitoring of end-effector movement.

Connected work zone devices also benefit from smart sensing technologies, including GPS, to enhance traffic monitoring and safety. In the broader context of smart traffic service systems, GPS serves as a linchpin, creating a connected environment. Integration with mobile devices and in-vehicle sensors enables the collection of real-time traffic data, forming the basis for intelligent decision support system presented in *(41)*. GPS receivers in this scenario created a pervasive, cost-effective means of gathering anonymized data crucial for effective traffic monitoring and planning within work zones. The use of GPS technology in a simplified architecture for connected work zone devices is further developed by *(35)*. These devices, equipped with GPS receivers, communicate data through a radio link to a central hub, enhancing work zone monitoring. This architecture reduces costs and simplifies data ownership and security concerns. GPS plays a crucial role in providing accurate location data, enabling effective communication and coordination among various devices within the work zone.

GPS trackers have also been employed in tracking workers on jobsites, specifically for assessing the effectiveness of Wearable Lighting Systems (WLSs) for worker visibility in highway construction projects. While nighttime road paving is a common procedure in order to reduce the traffic congestion during the daytime, it not only poses the risk of intrusion of vehicles into work zone but also increase the safety risk which resulted from the workers poorer visibility that can lead to injuries and fatalities. Evaluating the effectiveness of WLSs in enhancing workers vision, *(36)* used GPS trackers to provide a detailed understanding of worker movements during night-time paving and to enable a comprehensive analysis of the impact of WLSs on visibility. They recommended optimum locations for WLSs accordingly.

A persistent challenge in implementing GPS technology within work zones pertains to its accuracy. The attainment of high accuracy is crucial for the localization system to avert near-miss incidents and collisions in highway work zones. However, diverse studies have reported varying degrees of accuracy associated with the GPS sensors they employed. For instance, in the system outlined in *(34)*, which incorporated a wireless communication module alongside GPS, real-time equipment tracking achieved a precision of 20 meters in outdoor location tracking. In contrast, Lu et al. *(42)* documented an average error of less than 10 meters when combining GPS and dead reckoning technology with a Bluetooth beacon (situated on the roadside) to track a truck's location in a densely populated urban area *(43)* recorded an average error of 1.1 meters in an open area, escalating to 2.15 meters and 4.36 meters in scenarios with nearby obstacles while evaluating GPS for equipment tracking in an urban setting. Remarkably, *(44)* cited an accuracy of 10 centimeters for GPS. However, it is probable that they integrated GPS with additional sensors to achieve this level of precision, as standalone GPS systems typically have a positional error on the scale of meters. The accuracy challenge becomes more pronounced on uneven surfaces; nonetheless, GPS demonstrates enhanced precision when integrated with multiple sensors, as evidenced by various experimental studies *(45,46)*.

**Inertial Measurement Unit (IMU):** Originally designed for the inertial navigation systems of aircraft in GPS-denied environments, traditional IMUs have undergone a transformation with the development of microelectromechanical systems (MEMS) *(47)*. This advancement allowed for miniaturization, making it possible to attach IMUs to the human body. A typical wearable IMU comprises a MEMS accelerometer, gyroscope, and magnetometer. The accelerometer, utilizing a movable proof mass and mechanical dampers, measures changes in capacitance resulting from acceleration. The gyroscope employs a vibrating mechanical element to measure the Coriolis force caused by angular rates, while the magnetometer gauges the Earth's magnetic field through the deformation of the resonator caused by the Lorentz force generated from the interaction between electrical current and the external magnetic field *(48)*. The raw measurements from the wearable IMU enable the derivation of velocity, displacement, and orientation data. Initially embedded in smartphones or smart wristbands for cost-effectiveness, MEMS accelerometers or gyroscopes attached to wearable IMUs have been instrumental in monitoring kinematic movement.

The synergy of GPS and IMUs marks a significant leap in precision for localization applications, especially in dynamic environments like highway work zones. When combined, GPS and IMUs complement each other: GPS offers a broad positioning context, while IMUs fill in the gaps with detailed, high-frequency movement data. This combination results in a robust system capable of delivering precise location information even in challenging scenarios where GPS alone may falter. A notable example is the work of Wang et al., who developed a localization and proximity detection model using a GPS-aided inertial navigation system (INS) *(37)*. This system enhances real-time location information with additional data on heading and speed, captured through the IMU, thus enabling a more precise identification of hazardous situations. The GPS-aided INS employed in their study combines GPS receivers and an IMU to deliver accurate 3D position, speed, and orientation data at high update rates. The system's design features a state tracking module that gathers position, heading, and speed data, and a safety rules module that analyzes this data to pinpoint unsafe proximities. Controlled field experiments confirmed the system's localization accuracy at 0.7m, alongside a notable reduction in false alarms. Furthermore, the research presented in *(32)* demonstrates how accurate localization can be achieved using a wireless sensor network complemented by a high-accuracy localization algorithm. This system consists of a network of smart sensor cones equipped with GPS and IMU sensors. These sensor cones, including a master cone to map the topology of all the sensor cones, provide a comprehensive and precise overview of the work zone.

Although not a sensor, Wi-Fi technology has also been pivotal in enhancing real-time location tracking and communication in highway work zones. Wi-Fi, known for its high-speed wireless connectivity, is increasingly being integrated into Internet of Things (IoT) systems to facilitate efficient and real-time data exchange in dynamic environments like construction sites. A notable investigation into the performance of Wi-Fi-based IoT systems was conducted by Sabeti et al. *(49)*. This study focused on evaluating the latency of Wi-Fi based real-time communication between various devices within highway work zones. The IoT network in their case study incorporated diverse devices, including a camera, smart glasses, and a smartwatch, alongside standard networking and processing equipment. The research aimed to measure the communication delay in different network setups, particularly under conditions simulating a crowded work zone. The findings revealed that even in the most congested test scenario, the average latency was impressively low, at just 10 milliseconds.

**Vehicle Intrusion Detection and Alerting**

A significant safety concern in highway work zones arises from vehicles accidentally intruding into work zones from outside. Effectively detecting such incidents and implementing the necessary safety measures is crucial for enhancing overall work zone safety *(50–53)*. Identifying an intruding vehicle accurately is a complex task that involves considering multiple factors, such as the vehicle's speed and orientation. Several methods were proposed for the detection of intrusion hazards and subsequently alerting drivers and nearby workers *(54–64)*. A summary of previous studies on the application of sensor technologies for vehicle intrusion detection and alerting is provided in Table 3.

**Table 3.** Summary of key publications on the application of sensor technologies for intrusion detection and alerting

| Publication | Sensor technologies used | Proposed system | Application |
|---|---|---|---|
| Nnaji et al. *(65)* | Radar, GPS | Advanced Warning and Risk Evasion (AWARE) | Detects intrusion incidents and alerts drivers and workers. |
| Lin et al. *(66)* | Radar, camera | Active Work Zone Awareness Device (AWAD) | Collects vehicle speed and driver behavior data and gives notification for a potential intrusion hazard. |
| Chai et al. *(67)* | Infrared laser | An integrated control system for traffic safety | Measures speed and detects intrusion to ensure the safety of both drivers and workers. |
| Vorhes et al. *(35)* | LiDAR, GPS, accelerometer, gyroscope | A system of low-cost connected work zone devices | Detects traffic and intrusion hazards through a system of connected sensors and communicates data to a central server. |
| Martin et al. *(68)* | Ultrasound | An intrusion alarm system based on Wireless Sensor Network (WSN) | Detects vehicle intrusion through ultrasonic beams and sends alert to workers. |
| Phanomchoeng et al. *(69)* | Ultrasound | An ultrasound-based parameter arrays for long-distance auditory warning | Sends directional sound warning to potentially intruding vehicles without disturbing other vehicles. |
| Awolusi et al. *(7)* | Infrared | Safety line | Comprises a transmitter, receiver, and alarm unit, offering alerts to workers upon vehicle intrusion into the work zone. |

**Radar:** Radar sensor systems have been extensively used in highway safety research and practice to collect vehicle speed data *(70–75)*. In recent developments, radar sensors have expanded their role in smart work zone systems dedicated to enhancing highway work zone safety through detection of vehicle intrusion hazards. The Advanced Warning and Risk Evasion (AWARE) system exemplifies this expansion, employing a vicinity monitoring unit (VMU) equipped with GPS and radar systems to monitor specific environments. The primary goal of the AWARE system is to promptly alert workers within a work zone by generating visual and audio warnings in the event of an intrusion. Additionally, a personal warning unit ensures that workers within a projected impact area receive timely notifications *(65)*. Moreover, the application of radar sensors in work zone safety has evolved further. In a recent Active Work Zone Awareness Device (AWAD) experiment in Florida, a radar device with dynamic speed feedback LED signs was incorporated, aiming at collecting precise vehicle speed data to enhance overall safety measures within work zones *(66)*. These advancements underscore the continued integration of radar sensors in cutting-edge solutions for highway work zone management and safety.

The Wavetronix SmartSensor stands out as the preferred choice for speed detection radar sensors in various field experiments *(66,70,73,74,76)*. Specifically, studies have utilized the SmartSensor HD Model 125 to gauge vehicle speeds in proximity to work zone areas. This device is a radar-based system which offers a range of information, encompassing vehicle speed, vehicle counts, and average speed *(77)*. In addition to its functionality, the widespread adoption of this sensor device in research experiments may be attributed to its availability within state DOTs. For example, the Iowa DOT reportedly maintains over 500 Wavetronix sensors across the state for traffic data collection *(70)*. However, despite its prevalent use in research experiments, the literature falls short in offering a detailed exploration or evaluation of the advantages and limitations associated with this specific radar-based sensor device. While its functionality has been leveraged effectively, there remains a gap in understanding how its features align with varying experimental needs and conditions. A critical analysis of the Wavetronix SmartSensor's performance in different scenarios and under diverse environmental factors could provide valuable insights for researchers seeking the most suitable radar sensor for their specific applications. Additionally, an exploration of potential drawbacks or challenges associated with it would contribute to a more profound understanding of its utility and inform future advancements in radar-based sensing technologies.

Radar boasts advantages such as affordability, dependable performance in diverse weather conditions and dusty environments, and the reliable detection of substantial objects like other vehicles or people *(78)*. Despite these merits, one challenge lies in the occurrence of false alarms, where the operator may already be aware of an object behind the equipment that poses no danger. While these alarms may seem harmless, overreliance on them can be hazardous, as the operator might assume there is no genuine threat and proceed to reverse without a full understanding of the actual situation *(78–80)*. Moreover, within the array of features and parameters offered by radar sensors, the pivotal attributes of non-intrusiveness and ease of installation have significantly contributed to its emergence as a promising technology in highway environments *(81)*.

**Laser:** Laser sensors show great promise for improving safety in highway construction work zones. In on-site detection and warning systems, laser sensors are crucial. For example, a comprehensive traffic safety system has been developed by The Research Institute of Highway at the Ministry of Transport of China. This system integrates various technologies like wide-area network communication, short-range microwave communication, vehicle intrusion detection, synchronized flicker, and infrared laser speed measurement. It focuses on controlling vehicle speed in advance warning areas, providing synchronized warnings and intrusion detection in transition and buffer areas, and implementing multi-functional alarms in work areas. The system also includes synchronized warnings in the work area and the construction management center, along with remote video monitoring *(67)*.

LiDAR, a type of laser sensor, works by emitting laser light towards objects and analyzing the reflected light to measure distances and create detailed 3D images, enabling effective detection and tracking *(82)*. LiDAR stands out as a key component in Smart Work Zone (SWZ) safety systems. Its ability to create detailed 3D images regardless of lighting conditions makes it valuable for predicting and warning workers about potential vehicle intrusions *(83)*. Despite challenges like limited tracking ranges and sensitivity to adverse weather, LiDAR's accuracy and reliability make it vital for effective smart work zone (SWZ) solutions. As LiDAR technology advances and becomes more cost-effective, its integration into safety systems is increasingly feasible, opening the door to innovative solutions *(83)*.

The combination of laser sensors, including LiDAR, with connected work zone devices enhances safety measures. The architecture outlined in Reference *(35)* envisions a network of affordable monitoring and incident detection devices communicating via a radio link to a central hub. Equipped with LiDAR and other sensors, this hub forwards information to a central server, enabling real-time monitoring and incident response. LiDAR's role includes providing detailed 3D information, improving the detection of traffic-related elements such as cones and lane markings. Reference *(82)* emphasizes LiDAR's effectiveness in detecting TCDs, highlighting its robustness compared to camera-based methods. The utilization of device retro-reflectivity in LiDAR technology improves the robustness of detection, particularly in situations where lighting conditions may vary. The capacity of LiDAR to offer reliable and comprehensive three-dimensional data enhances its versatility as a tool for the detection and monitoring of lane markings. This underscores its potential in the challenging context of highway construction work zones.

Laser sensors, especially when incorporated into highway work zone monitoring systems, bring both advantages and face specific challenges. The requirement for real-time mapping in dynamic highway work zones poses challenges related to line-of-sight dependence, necessitating sensor repositioning when there is obstruction. The limitations of mapping sensors like laser scanners, relying on dynamic platforms, add complexity to the process *(84)*. However, recent technological advancements have addressed computational costs and processing times, making real-time applications more practical. The benefits of laser sensors, exemplified by LiDAR, are apparent in their ability to provide precise 3D data of highway work zones in real-time. Their high precision, quick response times, and extended detection range make them capable of delivering accurate and timely information. As seen in the findings from reference *(83)*, the LiDAR-based intrusion detection system showed an impressive precision rate of 100% in detecting vehicles. Despite these capabilities, laser sensors have challenges. Harsh weather conditions like heavy rain or fog can impact their performance, requiring additional measures for sustained effectiveness. Also, their sensitivity to interference from other light sources calls for careful deployment planning to enhance reliability. Moreover, interpreting data remains challenging due to the complex nature of unstructured point clouds and varying density, calling for ongoing improvements in algorithms and machine learning for effective scene understanding *(85)*. Important considerations include adapting sensor systems to changing highway work zone conditions, addressing privacy concerns, and ensuring transparent deployment.

**Ultrasonic Sensors:** Ultrasonic sensors utilize sound waves beyond the range of human hearing (ultrasonic waves) to detect and measure the distance to objects *(86)*. These sensors vary widely in their measurement ranges: short-range ultrasonic sensors can detect objects from a few centimeters up to 2-3 meters, while the range of long-range ultrasonic sensors can extend to more than 10 meters *(87)*. In *(68)*, ultrasonic sensors were employed for intrusion detection, where they formed a virtual barrier to identify when vehicles breached the perimeter of work zones. Ultrasound-based parameter arrays were investigated in *(69)*, for the possible application of generating a directional sound for long distance auditory warning of potential intruding vehicles, without disturbing other vehicles. Although ultrasonic sensors offer the advantage of tracking the speed of objects, surpassing other sensors like infrared, a significant drawback is their susceptibility to interference from noise or atmospheric conditions such as fog, dust, or rain. Therefore, they are generally considered more appropriate for indoor rather than outdoor applications *(88)*. Furthermore, their cost is taken to be relatively higher than other sensors for similar applications *(69,89)*.

**Infrared (IR) sensors:** IR sensors are a type of sensor that utilizes infrared radiation to detect and measure various physical parameters. Among them, Passive Infrared (PIR) sensors, a distinct subset, specialize in detecting infrared radiation, contrasting with active IR sensors that both emit and receive IR radiation. Their capability to detect motion, coupled with their effectiveness in low-light conditions, quick response time and low energy consumption, renders them as promising alternatives for highway intrusion detection applications *(86)*. An intrusion detection and alarming system based on IR sensors was investigated in *(7)*. The system consisted of an IR transmitter and receiver, combined with a mechanism which warns workers when a vehicle intrudes a work zone. In addition, PIR sensors are particularly utilized in other highway and traffic applications including vehicle speed estimation and queue warning system, which although the application is different the mechanism could be utilized for developing intrusion systems *(67,89)*.

**Worker-Equipment Proximity**

In addition to the external threat of vehicles entering work zones, a significant internal risk exists from workers coming into close proximity with construction equipment within work zones. Sensor technologies have been identified as effective tools for detecting and providing alerts in such scenarios, offering promising solutions to enhance safety measures *(90,91)*. Details of how sensor technologies can address these safety concerns are presented in the following sections, with a summary provided in Table 4.

**Table 4.** Summary of publications on the application of sensor technologies for proximity detection and warning

| Publication | Sensor technologies used | Proposed system | Application |
|---|---|---|---|
| Park et al. *(92)* | Bluetooth | Bluetooth based proximity detection and alert system | Utilizes Bluetooth to communicate in real-time and provide alerts to workers in hazardous situations. |
| Park et al. *(93)* | Bluetooth Low Energy (BLE) | Direction aware BLE-based proximity detection system | Detects proximity and direction of the incident and sends alerts to both workers and operators. |
| Kim et al. *(94)* | BLE | Internet of Things (IoT) based work zone proximity safety system | Consists of a network of sensor devices and logic for signal processing and alerting that detects incidents and timely activates alerting modules. |
| Teizer *(95)* | RFID | Self-Monitoring Alert and Reporting Technology for Hazard Avoidance and Training (SmartHat) | Utilizes battery-free passive RFID tags for detection and alerting of proximity incidents between workers and equipment. |
| Choe et al. *(87)* | Ultrasound, pulsed radar | Evaluation tests on sensing technologies for the prevention of back over accidents | Ultrasonic and radar-based systems were tested for their performance in detecting back over incidents across various scenarios. |

**Bluetooth and Bluetooth Low Energy (BLE):** Among other technologies, Bluetooth and Wi-Fi have become by far the most popular communication technologies used for various applications *(96)*. Bluetooth is a wireless technology standard designed for exchanging data over short distances using short-wavelength UHF radio waves *(97)*. Emerging from its conventional role in consumer electronics, Bluetooth technology has now been adapted to address the unique challenges of ensuring safety in highway work zones. A notable application of Bluetooth in this field is illustrated by Park et al. *(92)*, who developed a proximity warning and alert system using Bluetooth technology. This system incorporated an Equipment Protection Unit (EPU) installed on construction equipment and Personal Protection Units (PPUs) for ground workers and equipment operators. Experimentation and comparison with RFID and magnetic field-based systems highlighted the unique strengths of Bluetooth technology, despite magnetic sensing systems showing superior performance in certain aspects.

Bluetooth Low Energy (BLE), on the other hand has gained popularity over conventional Bluetooth technology due to mainly its efficient power consumption *(98)*. Introduced in 2010 by the Bluetooth Special Interest Group (SIG) as part of the Bluetooth 4.0 specification, it operates in the 2.5 GHz industrial, scientific, and medical (ISM) band *(93)*. BLE, compared to conventional Bluetooth, boasts attributes such as low cost, low energy consumption, ease of use, small form factor, and several other advantages *(98)*. BLE was employed in *(93)* to detect and alert proximity situations between ground workers and equipment. In this setup, location broadcasters – BLE sensors – were attached to construction equipment and personal protection units (PPUs). While this system demonstrated its utility in alerting equipment operators of potential hazards, it was observed to have a significant delay in signal reception, pointing to areas for further refinement. The use of signal processing algorithm proved to be effective in improving issues related to delay in signal reception and signal noise *(98)*.

Bluetooth and BLE are well-suited for short-range data communication at a local level. To extend their reach for remote connectivity, these systems are often integrated with the internet via additional technologies. This integration is exemplified in the IoT-based proximity safety system for work zones discussed in *(94)*. The system is designed to utilize Wi-Fi or 4G/5G networks for remote data storage and management, demonstrating a seamless integration of local and remote communication capabilities. It features a two-way data communication system, which also enables remote monitoring and allows personnel to adjust system settings and parameters remotely.

Bluetooth and BLE have several key strengths which make it suitable for proximity related highway work zone safety issues. Their rapid connectivity, cost-effective hardware, and minimal infrastructure requirements are particularly advantageous for such applications *(92)*. In addition, BLE offers a highly efficient energy usage capable of operating with a coin sized battery for years *(93)*.

**Radio Frequency Identification (RFID):** RFID is a sensor technology extensively used in proximity hazard detection and alerting systems. It operates by automatically identifying and tracking objects or people through small electronic devices, known as RFID tags. These tags can be read from a distance using RFID readers that emit radio waves to communicate with them *(99)*. RFID tags come in two types: active, which have their own battery for power, and passive, which draw power from the reader's signal during communication *(100)*. The technology primarily utilizes four frequency bands: low frequency (125/134 kHz), high frequency (13.56 MHz), ultra-high frequency (860 − 960 MHz), and microwaves (2.4 GHz) *(100)*. Each frequency band has its specific use and range, making RFID a versatile choice.

A passive RFID tag was used in *(95)* to develop a battery-free real-time proximity warning system named Self-Monitoring Alert and Reporting Technology for Hazard Avoidance and Training (SmartHat). The study also demonstrated through a field test that the system produced reliable results. However, one major drawback of passive RFID tags is their limited range; these tags have a range of only a few centimeters to about 15 meters *(44,95)*. In contrast, active RFID tags offer a more substantial range, up to 150 meters, as discussed in *(44,101)*. This significant difference in range could be a crucial factor when deciding between active and passive tags for various work zone scenarios.

RFID technology offers the benefit of creating sensor networks that are cost-effective, energy-efficient, and environmentally friendly *(100)*. The advantage of passive RFID tags lies in their independence from power sources, although they are limited by a short operational range. Conversely, active tags can function over longer distances and, despite relying on batteries, are noted for their efficiency, often running for years on a single battery *(102)*. However, the primary downside of active tags is their price, which is notably higher than that of passive tags *(102,103)*.

**Ultra-Wideband (UWB):** UWB technology is a wireless communication method that employs a wide range of frequencies to transfer data over short distances. UWB technology, with its exceptional accuracy ranging from 10 to 50 cm, proves to be tailor-made for highway construction work zones. UWB's latency of less than 1 ms makes it ideal for real-time location tracking in highway work zones *(104)*. The real-time location tracking capabilities of UWB was employed in *(105)* to create a connected work zone for automated detection of both proximity incidents and intrusion of Connected/Automated Vehicles (CAVs). The study, through tests conducted within smart road infrastructure, confirmed the effectiveness of the developed system. However, it also highlighted potential limitations in practical scenarios where sensing interruptions might arise. A significant advantage of UWB technology is its precision, capable of delivering accuracy to the centimeter level *(105)*. Nevertheless, when compared to other sensor technologies, UWB's comparatively higher cost can be a deterrent in its adoption for these applications.

Other sensors explored for proximity applications in work zones include ultrasonic and pulsed radar technologies. Choe et al. conducted an experimental evaluation of four commercially available sensor systems: one based on ultrasonic technology and three using pulsed radar technology *(87)*. In their study, the sensors were installed at the back of trucks, and their performance was assessed in relation to detecting objects, particularly workers. The system was specifically designed to assess construction back-over safety practices, a goal achieved through a series of tests. These included sensor installation review, static and dynamic testing, and a dirty sensor test to replicate real-world conditions.

**CHALLENGES AND FUTURE DIRECTIONS FOR THE IMPLEMENTATION OF SENSOR TECHNOLOGIES IN HIGHWAY WORK ZONES**

Highway work zones present unique challenges compared to general construction sites, including the presence of nearby live traffic, amplified noise levels due to moving vehicles, and their temporary nature with frequent shifts along road segments. Additionally, these zones often lack direct power supply and have confined operational areas to maintain traffic flow. As illustrated in Figure 4, a typical work zone layout comprises a transition area that redirects traffic away from the usual path, a dedicated working space for construction activities, and a surrounding buffer zone to safeguard the work area *(106)*. This layout is common for stationary temporary work zones, though variations like inclusion of lane shifts may occur. This environment demands sensor technologies that are not only precise in data collection, but also robust and adaptable enough to cope with the dynamic conditions *(72,93,98,107–109)*. Sensors are expected to be cost-effective for such temporary settings, power-efficient, capable of identifying all potential hazards, and durable enough to withstand conditions like adverse weather and constant vibrations. On the other hand, the successful application of these technologies hinges on their ability to accurately and promptly detect potential hazards, enabling timely interventions to prevent accidents and ensure worker safety. Hazardous situations can arise suddenly, as vehicles often move at high speeds, reducing the reaction time available to workers and drivers. Research shows that key factors in work zone crashes are the speed of the approaching vehicle and the sensor system's accuracy in detecting intrusions and issuing alerts *(52)*. Immediate incident detection and alerting are crucial in these scenarios.

Furthermore, the deployment of sensor technologies in such a setting must take into account the user-friendliness aspect. Given the diverse technical skills of the workforce in highway work zones, it is essential that these devices are intuitive and easy to use, minimizing the learning curve and facilitating seamless integration into daily operations. Other key considerations include economic feasibility, durability, security, and adherence to regulations. This section aims to delve deeper into these specific challenges and characteristics of highway work zones, exploring how they shape the required performance of sensor technologies. It also provides insights and recommendations on the future prospects of these technologies and best practices for their implementation. Before these sections, a summary of the comparative advantages and disadvantages of different sensor technologies is provided in Table 5. This comparison is not intended to provide precise data on the various aspects of these technologies, given their wide range of properties influenced by numerous factors. Instead, it aims to offer general insights into the key characteristics that distinguish these technologies from one another.

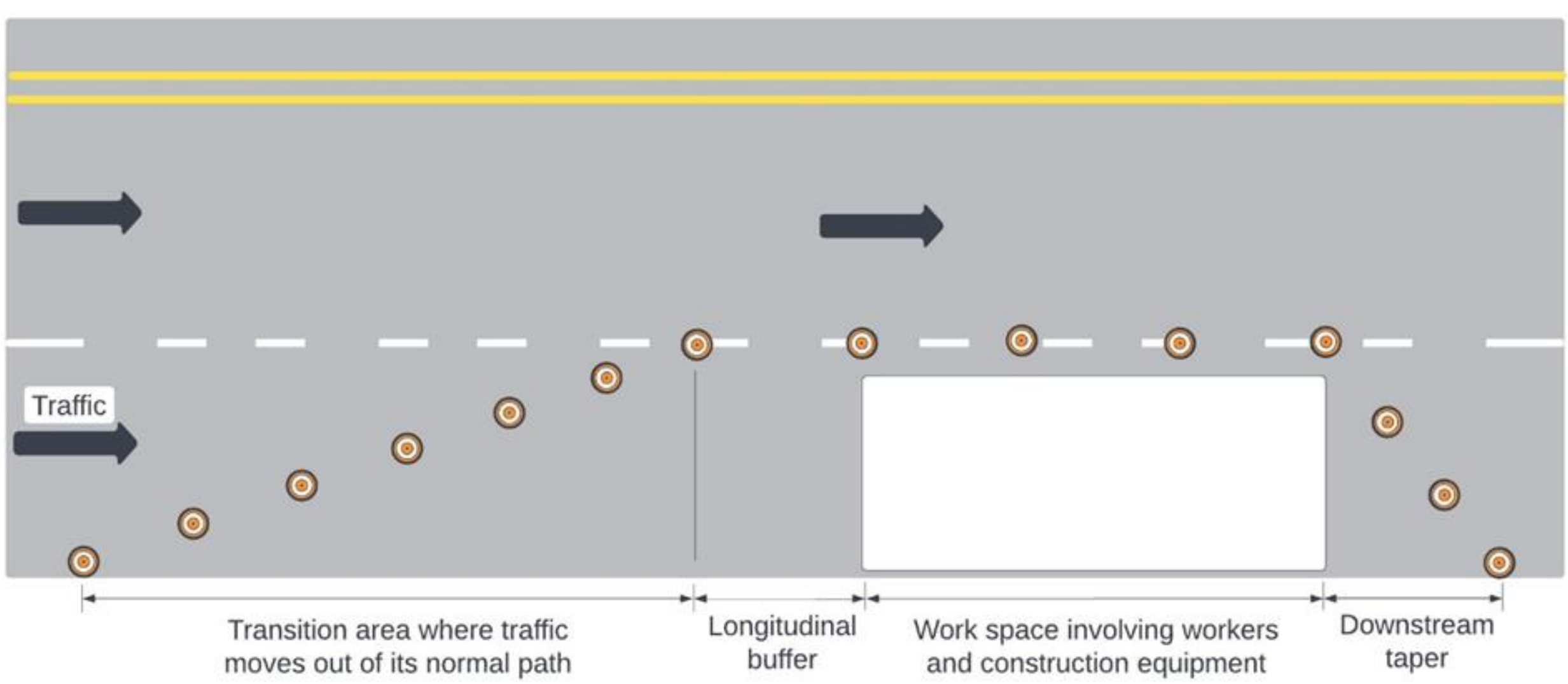

**Figure 4.** Typical highway work zone set-up

**Table 5.** Summary of performance comparison between different sensor technologies

| Sensors | Advantages | Disadvantages | Relative cost |
|---|---|---|---|
| GPS | Global coverage. Relatively accurate over long distances | Low accuracy; effective only in outdoor situations; accuracy is affected by obstacles | Low to medium |
| IMU | Good for orientation and movement tracking; works indoors and outdoors; complements well with GPS | Drift over time; requires initial calibration. | Low to medium |
| Radar | Dependable performance in diverse weather conditions and dusty environments; non-intrusive and easy to use | False alarms during usage for intrusion detection | Medium to high |
| Laser | Laser technologies like LiDAR can provide precise 3D map of work zones in real-time | Dependent on line-of-sight; can be affected by harsh weather conditions like heavy rain or fog; complexity in interpreting data | High |
| Ultrasonic | Good for close range measurements | They are not preferred for outdoor applications, as they are affected by noise or environmental conditions like fog, dust, and rain | Low; may be relatively higher than some other sensors |
| IR | Simple to use; can work in dark conditions | Limited range; can be affected by sunlight and other IR sources | Low |
| Bluetooth/BLE | Low energy consumption; good for short-range communication | Limited range; security concerns | Low |
| RFID | Energy efficient and environmentally friendly | Passive RFID tags have relatively short range | Low to medium (active tags have higher cost than passive ones) |
| UWB | High accuracy for distance and positioning; works well indoors | Requires more power than some alternatives | Medium to high |

**Size and Weight**

As sensors come in different sizes and weights, choosing the one with the appropriate size will be an important task. Sensors in highway work zone environments are generally preferred to have small size and weight so that it will be convenient to place on workers, equipment, traffic cones, and other facilities. In an environment such as highway work zones, where space is limited, the use of small and compact sensors will also help for a convenient workplace and avoids obstructions. Although the size and weight of sensor devices could depend on many other factors including the manufacturer and the specific model, the sensor technology that they are using is also a key factor.

Many sensor technologies including GPS, RFID, Bluetooth, BLE, Infrared and IMU often come with compact sizes convenient for highway work zone applications *(93,110)*. However, comparing between the different technologies, some may be even smaller than others. For instance, passive RFID tags, benefiting from their battery-less design, are notably smaller than their active counterparts *(100)*. This compactness makes them particularly convenient for integration into equipment or wearable items like worker badges and vests. However, it is important to note that passive RFID tags, despite their size advantage, may not be ideal for certain work zone safety applications. They lack critical features such as real-time location tracking, long-range capabilities, and omnidirectional communication with the RFID reader, which are essential in dynamic work zone environments *(102,103)*. On the other hand, sensors like Bluetooth, BLE, and IMUs usually come embedded in common devices such as smartphones and smartwatches *(111)*.

A related issue is the implementation of modular design, which involves dividing a system into smaller, self-contained elements or modules. Modular design is crucial in ensuring adaptability, expansion, and integration simplicity *(112)*. Due to the temporary and dynamic character of work zones, there is significant variation in dimensions, arrangement, and operational structure. The lack of attention to modular design principles restricts the ability to customize and adapt safety systems to meet the unique demands of each individual work zone. The implementation of modular systems would enable safety engineers to construct safety solutions that are specifically designed for each work zone, thereby promoting the development of more efficient and situation-appropriate safety protocols. Instead of building a completely new system for each work site, the modular design offers the ability to modify the sensor modules based on new site properties and demands.

Moreover, the rapid development of sensor technology requires systems that can easily incorporate new improvements. Existing safety systems encounter integration difficulties with developing technologies in the absence of a modular design approach *(113)*. Integrating new sensors or communication protocols into non-modular systems becomes challenging and expensive. Implementing a modular framework would accelerate the easy integration of state-of-the-art technology, which ensures that the system takes advantage of the newly invented sensors. The lack of modularity presents difficulties in terms of the capacity to upgrade and maintain the system. Additionally, from repair and maintenance point of view, in non-modular systems, upgrading a single component may necessitate significant overhauls or replacements of the entire system. Modular designs, on the other hand, enable targeted upgrades, reducing downtime, and minimizing costs associated with maintenance and system enhancements.

Modular design can be implemented by developing a standardized interface protocol for sensor modules, facilitating plug-and-play functionality for various sensors and standardizing the communication interface. This approach enables the quick and efficient interchange of sensor modules based on specific work zone requirements, eliminating compatibility issues. Additionally, developing a universal mounting system for these sensor modules can further enhance their adaptability in diverse work zone scenarios. Such a system should be designed for easy attachment to a wide range of surfaces and objects commonly found in work zones, including construction equipment, traffic cones, and worker devices. A universal mounting system ensures that regardless of the sensor type or manufacturer, modules can be securely and effectively integrated into the work zone environment. This method streamlines the process of deploying sensor technologies and significantly reduces the time and resources needed for installation and reconfiguration.

Overall, the lack of focus on modular design principles in current literature presents significant barriers to the development, flexible expansion, and customization of highway work zone safety systems. For the purpose of advancing safety technologies in dynamic work zone contexts, future research should prioritize the development and study of modular structures. This will allow for the identification and resolution of the issues that have been identified, as well as the utilization of the benefits that modularity offers.

**Accuracy and Range**

In the context of safety in highway work zones, accurately identifying intrusion and proximity hazards is crucial and hinges on precise measurement of location, speed, and orientation, coupled with real-time notification capabilities. This necessitates sensors with high accuracy, a key criterion in their selection. While the practice of sensor fusion, which involves combining data from multiple sensors to enhance accuracy, is prevalent, the inherent accuracy of each individual sensor remains a fundamental aspect. This is because the overall effectiveness of sensor fusion largely depends on the individual accuracy of the sensors used in the process. On the other hand, when considering sensors for highway work zones, it is essential that they possess a sensing range that is adequately wide to ensure comprehensive coverage of the significant areas of the work zone. Achieving such coverage may often involve the deployment of sensor clusters. However, the importance of the sensing range becomes particularly pronounced when the use of sensors is limited in number. Taking into account the dimensions of work zones and the distance expected to be covered, these sensors can be broadly classified into three categories: long-range, medium-range, and short-range *(99)*. Long-range sensors, such as GPS, radar, and Ultra-Wideband (UWB), are known for their extensive coverage capabilities, making them suitable for wide-area monitoring encompassing the work zone and its surroundings. Medium-range sensors, which include technologies like laser, and Radio-Frequency Identification (RFID), offer a balance between range and precision. Finally, sensors that fall into the short-range category, including infrared, Bluetooth, Inertial Measurement Units (IMUs), and ultrasonic sensors, are tailored for more localized applications within smaller areas *(97,99,110)*.

The accuracy and range of sensor readings in highway work zones can be significantly enhanced through the implementation of sensor fusion algorithms. These algorithms apply sophisticated methods that integrate data from various sensors, such as GPS and IMUs, to produce more comprehensive and accurate results *(63)*. By intelligently combining the strengths of each sensor type, sensor fusion compensates for the individual limitations inherent in each sensor. This integrated approach is particularly effective in dynamic and complex work zone environments, where single-sensor data might be insufficient or prone to inaccuracies due to environmental factors like obstructions. In practice, sensor fusion algorithms can enhance the precision of location tracking, speed measurement, and orientation detection in real-time, enabling more reliable hazard identification and decision-making processes *(63)*.

**Power Source and Consumption**

Many highway work zones are often situated in remote or undeveloped areas where access to a reliable and convenient power supply is limited. This scarcity is primarily due to the transient nature of these zones and the logistical challenges in establishing a permanent power infrastructure in areas that are not typically served by the power grid. In addition, the fluctuating and short-term usage of these zones makes it impractical to invest in extensive power delivery systems. Therefore, sensors deployed in such environments are required to be power-efficient and adaptable in terms of power sourcing. This necessity, however, introduces a tradeoff between the size and weight of sensors and their power efficiency, as higher efficiency often necessitates larger and heavier batteries. In this context, BLE sensors are a notable example of power-efficient technology. They are designed to consume significantly less energy than traditional Bluetooth devices, enabling them to operate on small batteries for extended periods, often lasting for years *(93,114)*. This long-term operation without frequent battery replacements makes them particularly suitable for use in work zones where power sources are scarce. RFID technology, particularly passive RFID tags, offers another power-efficient solution. Passive tags require no internal power source, deriving their energy from the reader's signal *(102)*. On the other hand, active RFID tags, though requiring a power source, are equipped with batteries that can last for several years *(102)*.

It is also important to note that the power consumption of these sensors can significantly vary depending on user-configured settings *(97)*. Settings that require higher data transmission rates, increased operational frequency, or extended range can lead to increased power usage. Therefore, in environments like highway work zones where power availability is a prime concern, it is crucial to strategically optimize these settings. This balance between performance and power conservation must be carefully managed to ensure both the efficacy and longevity of the sensor systems. Such considerations have led to the development of sensor systems specifically designed for low power consumption, ensuring prolonged operational periods. These considerations have been pivotal in the design of several existing sensor systems tailored for work zones *(35,37,95)*. Furthermore, strategies like employing voltage regulators and integrating a low-power or 'sleep' mode have been proposed to further curtail power consumption *(35)*. In the future, the use of readily available power sources like solar energy could offer a more reliable and sustainable alternative for powering sensor systems in highway work zones *(35)*.

**Cost Implications**

Affordability and practicality in costs are essential when deploying sensors in temporary settings like highway work zones, where extensive investment in sensor technology might not be justifiable. The expenditure on these sensors should be proportional to the benefits and functionalities they provide. The typical cost of existing smart work zone devices is known to be between 1% to 5% of the total project budget *(35)*. Factors influencing the cost include the price of individual sensors, the quantity required to adequately monitor the work zone, the infrastructure necessary for sensor integration, and operational cost *(115)*. While cost-effectiveness is a critical consideration in sensor selection, it is crucial to balance this with the quality and suitability of the sensors for specific work zone needs. For instance, passive RFID tags, known for their cost-efficiency compared to active RFID tags, might fall short in applications requiring real-time location tracking of workers and equipment in highway work zones *(102,103)*. In a broader view, technologies like BLE, infrared, RFID, and ultrasound generally present more economical options *(93,110)*. However, it is worth noting that within these categories, there can be cost variations; for example, ultrasonic sensors tend to be more expensive than their passive infrared counterparts *(86,89)*.

The findings of this review indicate that the financial criteria have been significantly neglected in the existing literature proposing sensor-based work zone safety improvement systems. Although there are a few studies that propose decision-making frameworks for selecting safety technologies in highway construction, these studies often fall short in adequately addressing the financial implications of using different systems. For example, a five-step decision-making framework utilizing the Choosing by Advantages (CBA) method was suggested in *(115)*. The defined criteria of this framework include the identification of general and in-category objectives, performance assessment of available technologies, and the selection and implication of the appropriate technology. However, in a case study conducted to illustrate the implications of the framework, cost and financial aspects were not included among the criteria for selecting alternative technologies. One possible explanation for the absence of cost-related information in the field may be the recent attention and need for developing such systems, which has led to prioritizing performance criteria over financial considerations. However, this emphasis on performance metrics without a parallel focus on financial implications presents a significant gap in the decision-making process, potentially hindering informed choices regarding the adoption and implementation of these technologies. To ensure a comprehensive evaluation and facilitate effective decision-making, future studies must strive to strike a balance between performance and implication costs within decision-making frameworks.

Implication costs, including but not limited to item price, the life cycle period, maintenance and replacement, and the impact of the system on worker productivity, are of even greater importance when it comes to sensor-based systems. This is because these systems often involve sophisticated technologies, requiring substantial initial investments that might not be transparently addressed in the existing literature. The financial implications extend beyond the mere acquisition cost of the sensor technologies, encompassing their entire life cycle. The lack of comprehensive financial assessments in the literature leaves a critical gap in understanding the overall cost-effectiveness and feasibility of implementing these systems.

One key aspect that is frequently overlooked is the consideration of maintenance and replacement costs. Sensor-based systems, by their nature, may require regular maintenance to ensure accurate and reliable performance. The frequency and complexity of maintenance procedures, as well as the associated costs, should be thoroughly evaluated. Additionally, understanding the expected lifespan of these systems and the costs associated with periodic replacements is essential for long-term planning and financial forecasting. Furthermore, the impact of these systems on worker productivity is a vital yet overlooked facet in existing literature. The implementation of sensor-based safety systems has the potential to enhance overall work zone safety, but it may also introduce new dynamics affecting the efficiency and productivity of the workforce. For instance, the learning curve associated with adopting and adapting to these technologies, potential disruptions caused by maintenance activities, and the need for specialized training for workers can all influence productivity. The economic implications of these factors need careful consideration for a holistic evaluation of the cost-effectiveness of sensor-based safety solutions.

**Ease of Use and User Acceptance**

In application of sensors in practical work zone situations, the goal is to choose sensor technologies that enhance safety and efficiency without adding complexity to the work environment. The user-friendliness of sensors is a critical consideration, especially in the high-stress, fast-paced environment of highway work zones. According to a survey collected in *(65)*, ease of difficulty in operation and maintenance was pointed out as one of the top three barriers in adopting smart work zone safety technologies, along with false alarms and inadequate work zone coverage. In addition, it was used in *(115)* as one of the four criteria for evaluating intrusion detection and alert systems, along with cost, effectiveness of the alarm, and effectiveness of the triggering mechanism. Sensors should be straightforward to deploy, operate, and maintain, with minimal training required for workers. The ease of use is vital to ensure that the technology integrates seamlessly into the work zone without causing disruptions or requiring extensive technical support. The use of some previously developed systems was hindered by their difficulty in installation and usage *(7)*. This included a microwave-based alarm system, which was rejected by the Alabama, Colorado, Iowa, and Pennsylvania DOTs because of setup problems *(7)*. On the other hand, ease of deployment is mentioned to be one of the major advantages of wireless sensor networks (WSNs) *(68)*.

While some studies touch upon the challenges of workforce and drivers acceptance *(73,116)*, there is a noticeable absence of comprehensive models specifically designed to address user acceptance in the specific context of highway work zones. Limited attention has been devoted to understanding how operators, construction workers, and other personnel engage with these systems on a day-to-day basis. Insights into how sensor data informs decision-making, potential challenges faced by the workforce in adopting new technologies, and the impact on overall work dynamics are essential components that require thorough investigation. The exploration of the aspects that influence the acceptability of sensor technologies, such as perceived usefulness, simplicity of use, and impact on job performance, is an area that has not yet been well studied. Creating models that capture the complex interaction between technology features and human aspects can greatly enhance the successful integration of various systems. It is also recommended to overcome consumer, here mostly workers, resistance towards new technologies, it is crucial to develop comprehensive yet user-friendly instructions that clarify the device's functionality, data collection processes, and data security measures *(44)*.

A key consideration in the successful implementation of sensor technologies is the level of training provided to the workforce. When conducting experiments, studies usually assumed the familiarity of workforce with their proposed sensory system and its functionality. Although it is an accepted approach in academic research, the real-world applications of such systems always involve challenges regarding the difficulty of use and sufficient knowledge of workers. It is crucial to comprehend the learning process involved in adopting sensor technologies and the elements that either support or impede the integration of these instruments into regular routines. Moreover, when a wide range of publications can be found on the critical success factors and adoption barriers of construction new-emerged technologies such as Building Information Modeling (BIM) or Virtual Reality (VR), it is still a wide-open path in highway work zone safety systems which needs to be taken in future studies.

Except for these, one critical human related aspect that is a challenge for all fields and industries which have planned to adopt advanced technologies, including road infrastructure construction, is the psychological impact of constant surveillance and interaction with sensor technologies. While this seems a more tangible issue when it comes to vision-based sensory and inventory systems, the efficacy of other types of systems, specifically the ones comprising of sensors attached to workers, can also be affected by this human-centric factor. The issue of reliability and trust in online platforms and location-based services is an ongoing problem for users, which is constantly increasing *(44)*. The literature gap lies in understanding how the continuous monitoring and alerting features of these systems influence the stress levels, job satisfaction, and overall well-being of the workforce. Addressing these concerns is pivotal for creating a work environment that promotes both safety and the mental and emotional welfare of the individuals involved.

**Long-Term Performance and Durability**

One notable observation within the current body of literature is the limited exploration of extended performance metrics. While initial studies often focus on the immediate benefits and real-time performance criteria, there is a lack of information regarding how these sensor technologies tolerate the test of time and prolonged exposure to the challenging environmental conditions prevalent in highway work zones. Highway work zones are dynamic environments characterized by unpredictable weather patterns, varying temperatures, and constant exposure to debris and construction activities. Despite the acknowledgment of these challenges, a comprehensive understanding of how sensors endure real-world conditions and potential wear-and-tear remains notably absent. Again, due to the recent emergence of smart work zone sensor-based systems, the durability and long-term performance of such systems remained uncovered. This fact, although natural and rational, emphasizes the necessity of longitudinal studies tracking the performance of sensor technologies through seasons, construction phases and locations, and diverse weather events.

However, it should be noted that there might be some overlap between durability and cost analysis of systems. In other words, when estimating the maintenance and replacement costs of elements of such systems, their long-term performance and durability in various geographical, environmental and weather conditions should be taken into consideration. Therefore, the knowledge gap in long-term performance of sensor-based highway work zone safety improvement systems could also affect the cost-effectiveness analysis of their implementation. This includes considerations of maintenance costs, the frequency of replacements, the economic feasibility of sustained sensor deployment in highway work zones.

Besides, calibration drift, a phenomenon where sensors gradually deviate from their initial calibration settings, is a pertinent concern that necessitates investigation. Ensuring the accuracy and reliability of data over extended periods is crucial for making informed decisions and maintaining the effectiveness of safety measures. It is even more important when considering the delicate and dynamic nature of highway work zones, compared to static constructions such as buildings.

**Cyber Security**

Besides the privacy and ethical consideration of using continuous data collection sensors, one notable research gap lies in the inadequate coverage of data security measures within sensor-based systems. Cyber security and data privacy are consistently paramount concerns in situations involving wireless communication *(117)*, which forms the dominant context of work zone safety systems. While studies highlight the data collected for real-time monitoring and decision-making, there is a lack of comprehensive investigations about the encryption, storage, and transmission techniques used to protect this sensitive information. Furthermore, work zone sensory systems are susceptible to cyber-attacks, and a depth understanding of the potential threats, vulnerabilities, and countermeasures is needed to protect these systems. The vulnerability of sensor networks to cyber threats and unauthorized access necessitates a more robust exploration of cryptographic techniques and secure communication frameworks designed to meet the unique demands of highway work zones. Robust cybersecurity protocols, intrusion detection mechanisms, and contingency plans in the event of a breach demand thorough exploration to ensure the resilience and reliability of sensor networks and communication.

**Regulatory and Legal Considerations**

The issue of liability and accountability in the event of sensor system failures or accidents within work zones is a critical consideration that requires more attention, particularly when considering the high fatality rate of the highway construction accidents. By increasing the implication of smart sensor-based work zone safety improvement systems, the potential of encountering system failures, such as system malfunctions, false positives/negatives, or accidents will accordingly grow. Research conducted in this field should offer detailed understandings of the legal structures that determine responsibility and create systems for holding individuals accountable, so providing clear explanations to all parties involved.

Highway construction projects may involve cooperation between different agencies and authorities and cross multiple levels of government. The existing literature lacks in-depth analysis of interagency collaboration arrangements and the legal obstacles arising from jurisdictional barriers. The objective of the research should be to create models that facilitate smooth collaboration, while considering the legal complexities related to sensor installations across different authorities. Moreover, the rapid evolution of sensor technologies requires a dynamic regulatory framework that can adapt to emerging innovations. In the context of the literature reviewed for this study, there appears to be a lack of comprehensive discussion on how regulatory bodies might proactively adapt their frameworks to incorporate technological advancements, while also ensuring safety and legal compliance. Future research should explore mechanisms for dynamic regulatory adaptation to ensure that the legal landscape remains agile and responsive to evolving sensor technologies.

## CONCLUSION

This comprehensive review has systematically examined the application and impact of sensor technologies in preventing intrusion and proximity hazards in highway work zones. Through a detailed analysis of various sensor types, their performance characteristics, and specific roles in work zone environments, the study presents a holistic view of current technological advancements and their practical implications. The results of the review indicate a significant potential for sensor technologies to mitigate risks associated with highway work zones. Their ability to detect intrusion, proximity hazards, and accurately track the location of workers and equipment positions them as invaluable tools in the quest to improve work zone safety.

The review highlights the unique advantages of different sensor technologies for localization, intrusion detection, and managing worker-equipment proximity. Selecting the right technology is essential for maximizing safety benefits in work zones. The following points summarize the distinct sensor technologies and their respective advantages and disadvantages:

- The research identified GPS and IMU sensors as key technologies for accurately tracking the location of workers and equipment within work zones. GPS provides broad location data, while IMUs deliver precise motion tracking, allowing for accurate and real-time updates of positions and movements.
- For detecting vehicle intrusions into work zones, the study highlights the use of radar, LiDAR, ultrasonic, and infrared sensors. These technologies identify unauthorized or unexpected vehicle entries, preventing accidents and ensuring occupant safety. Radar and LiDAR provide broad, comprehensive data but are more costly. In contrast, ultrasonic and infrared sensors are more economical but have shorter range capabilities, making them ideal for localized applications.
- Addressing the proximity between workers and equipment, RFID, Bluetooth/BLE, ultrasonic, UWB, and infrared technologies were utilized in existing research. These sensors detect close-range hazards, enabling timely alerts to prevent collisions and injuries. Technologies such as RFID and BLE are energy-efficient and cost-effective but have limited range. UWB offers a wider detection range but comes with higher costs and power consumption, presenting a trade-off between range and efficiency.

To ensure the effective application of sensor technologies in highway work zone safety, several critical considerations must be taken into account. These key factors, outlined below, directly influence the practicality, efficiency, and success of deploying these technologies:

- Effective sensor deployment requires seamless integration with existing work zone management and safety systems. This compatibility ensures that sensor data can be effectively utilized without the need for extensive modifications to current processes.
- Considering the temporary nature of work zones, the cost of sensor technologies, both initial and operational, must be justified by the value they add in terms of enhanced safety and efficiency. This includes evaluating the total cost of ownership, from purchase through maintenance to eventual replacement.
- The design of sensor systems should take into account the ease of use for work zone personnel. This includes both the physical deployment of the sensors and the interface for monitoring and responding to the data they generate, ensuring that workers can effectively leverage these tools without extensive training.

Incorporating sensor technologies in highway work zones to enhance safety and efficiency brings forth a series of challenges that necessitate careful consideration and management. Among these, certain concerns stand out as particularly significant due to their potential impact on the successful deployment and acceptance of these technologies:

- The introduction of wearable sensors for tracking purposes raises concerns among workers regarding privacy and autonomy. Ensuring that these technologies are accepted requires clear communication about their purpose, benefits, and the measures in place to protect workers' privacy and personal data.
- As sensor technologies often collect and transmit sensitive information, robust cybersecurity protocols are essential to prevent unauthorized access and data breaches. Protecting the integrity and confidentiality of collected data is paramount to maintaining trust in the technology.
- Ensuring that sensor technologies comply with current regulations and standards is a significant concern. This includes navigating the evolving legal landscape concerning data privacy, usage, and the liability issues that may arise from sensor malfunctions or misinterpretations, which could lead to legal challenges.

Looking to the future, efforts should focus on developing modular sensor systems that are flexible and easy to upgrade, facilitating seamless adaptation to new challenges and simplifying maintenance processes. Comprehensive cost assessments are also essential, encompassing initial purchases, ongoing operational expenses, maintenance, and future upgrades to ensure informed decision-making. Addressing implementation barriers, such as technical limitations and user acceptance issues, is crucial for the successful deployment of these technologies. Additionally, refining regulatory and cybersecurity measures will be vital to ensure compliance with evolving legal standards and to protect against cyber threats.

**FUNDING**

The authors disclosed receipt of the following financial support for the research, authorship, and/or publication of this article: This research was supported by the Minnesota Department of Transportation (MnDOT), Grant No. 1036338.

**AUTHOR CONTRIBUTION STATEMENT**

The authors confirm contribution to the paper as follows: study conception and design: Ayenew Yihune Demeke, Moein Younesi Heravi, Israt Sharmin Dola, Youjin Jang, Chau Le, Inbae Jeong, Zhibin Lin, Danling Wang; data collection: Ayenew Yihune Demeke, Moein Younesi Heravi, Israt Sharmin Dola; analysis and interpretation of results: Ayenew Yihune Demeke, Moein Younesi Heravi, Israt Sharmin Dola, Youjin Jang, Chau Le, Inbae Jeong, Zhibin Lin, Danling Wang; draft manuscript preparation: Ayenew Yihune Demeke, Moein Younesi Heravi, Israt Sharmin Dola, Youjin Jang, Chau Le, Inbae Jeong, Zhibin Lin, Danling Wang. All authors reviewed the results and approved the final version of the manuscript.

**DECLARATION OF CONFLICTING INTERESTS**

The authors declare no potential conflicts of interest with respect to the research, authorship, and/or publication of this article.

**REFERENCES**

1. Nnaji, C., J. Gambatese, H. W. Lee, and F. Zhang. Improving construction work zone safety using technology: A systematic review of applicable technologies. *Journal of Traffic and Transportation Engineering (English Edition)* Preprint at https://doi.org/10.1016/j.jtte.2019.11.001 (2020) vol. 7.
2. Highway Work Zone Safety | NIOSH | CDC. https://www.cdc.gov/niosh/topics/highwayworkzones/default.html.
3. Sakhakarmi, S., and J. W. Park. Improved intrusion accident management using haptic signals in roadway work zone. *J Safety Res*, 2022. 80.
4. Work Zones - Injury Facts. https://injuryfacts.nsc.org/motor-vehicle/motor-vehicle-safety-issues/work-zones/ (2023).
5. Work Zone Safety Data | Bureau of Transportation Statistics. https://www.bts.gov/browse-statistical-products-and-data/national-transportation-statistics/work-zone-safety-data.
6. Yang, X., and N. Roofigari-Esfahan. Vibrotactile Alerting to Prevent Accidents in Highway Construction Work Zones: An Exploratory Study. *Sensors*, 2023. 23.
7. Awolusi, I., and E. D. Marks. Active work zone safety: Preventing accidents using intrusion sensing technologies. *Front Built Environ*, 2019. 5.
8. Sabeti, S., O. Shoghli, M. Baharani, and H. Tabkhi. Toward AI-enabled augmented reality to enhance the safety of highway work zones: Feasibility, requirements, and challenges. *Advanced Engineering Informatics*,

2021. 50.
9. Tremmel, M., U. G. Gerdtham, P. M. Nilsson, and S. Saha. Economic burden of obesity: A systematic literature review. *International Journal of Environmental Research and Public Health* Preprint at https://doi.org/10.3390/ijerph14040435 (2017) vol. 14.
10. Bogic, M., A. Njoku, and S. Priebe. Long-term mental health of war-refugees: a systematic literature review. *BMC Int Health Hum Rights*, 2015. 15.
11. Raes, A., L. Detienne, I. Windey, and F. Depaepe. A systematic literature review on synchronous hybrid learning: gaps identified. *Learning Environments Research* Preprint at https://doi.org/10.1007/s10984-019-09303-z (2020) vol. 23.
12. Durach, C. F., J. Kembro, and A. Wieland. A New Paradigm for Systematic Literature Reviews in Supply Chain Management. *Journal of Supply Chain Management*, 2017. 53.
13. Kraus, S., M. Breier, and S. Dasí-Rodríguez. The art of crafting a systematic literature review in entrepreneurship research. *International Entrepreneurship and Management Journal*, 2020. 16.
14. Aleti, A., B. Buhnova, L. Grunske, A. Koziolek, and I. Meedeniya. Software architecture optimization methods: A systematic literature review. *IEEE Transactions on Software Engineering*, 2013. 39.
15. Carvalho, T. P., *et al.* A systematic literature review of machine learning methods applied to predictive maintenance. *Comput Ind Eng*, 2019. 137.
16. Badi, S., and N. Murtagh. Green supply chain management in construction: A systematic literature review and future research agenda. *J Clean Prod*, 2019. 223.
17. Hijazi, A. A., S. Perera, R. N. Calheiros, and A. Alashwal. Rationale for the Integration of BIM and Blockchain for the Construction Supply Chain Data Delivery: A Systematic Literature Review and Validation through Focus Group. *J Constr Eng Manag*, 2021. 147.
18. Fink, A. G. *Conducting Research Literature Reviews: From The Internet To Paper*. (2005).
19. Moher, D., *et al.* Preferred reporting items for systematic reviews and meta-analyses: The PRISMA statement. *PLoS Medicine* Preprint at https://doi.org/10.1371/journal.pmed.1000097 (2009) vol. 6.
20. Moher, D., *et al.* Preferred reporting items for systematic review and meta-analysis protocols (PRISMA-P) 2015 statement. *Revista Espanola de Nutricion Humana y Dietetica*, 2016. 20.
21. Ranasinghe, U., M. Jefferies, P. Davis, and M. Pillay. buildings Conceptualising Project Uncertainty in the Context of Building Refurbishment Safety: A Systematic Review. 2021. doi:10.3390/buildings.
22. Chen, X., *et al.* Implementation of technologies in the construction industry: a systematic review. *Engineering, Construction and Architectural Management*, 2022. 29.
23. Babalola, A., P. Manu, C. Cheung, A. Yunusa-Kaltungo, and P. Bartolo. A systematic review of the application of immersive technologies for safety and health management in the construction sector. *J Safety Res*, 2023. 85.
24. Kim, S., and K. Kim. Systematic Tertiary Study for Consolidating further Implications of Unmanned Aircraft System Applications. *Journal of Management in Engineering*, 2021. 37.
25. Royle, P., N. B. Kandala, K. Barnard, and N. Waugh. Bibliometrics of systematic reviews: analysis of citation rates and journal impact factors. *Syst Rev*, 2013. 2.
26. McCrae, N., M. Blackstock, and E. Purssell. Eligibility criteria in systematic reviews: A methodological review. *International Journal of Nursing Studies* Preprint at https://doi.org/10.1016/j.ijnurstu.2015.02.002 (2015) vol. 52.
27. Nnaji, C., J. Gambatese, H. W. Lee, and F. Zhang. Improving construction work zone safety using technology: A systematic review of applicable technologies. *Journal of Traffic and Transportation Engineering (English Edition)* Preprint at https://doi.org/10.1016/j.jtte.2019.11.001 (2020) vol. 7.
28. Ayodele, O. A., A. Chang-Richards, and V. González. Factors Affecting Workforce Turnover in the Construction Sector: A Systematic Review. *J Constr Eng Manag*, 2020. 146.
29. Ahmad, S., *et al.* Evaluating journal quality : A review of journal citation indicators and ranking in library and information science core journals. *COLLNET Journal of Scientometrics and Information Management*, 2019. 13.
30. Pajić, D. On the stability of citation-based journal rankings. *J Informetr*, 2015. 9.
31. Li, H., G. Chan, J. K. W. Wong, and M. Skitmore. Real-time locating systems applications in construction. *Autom Constr*, 2016. 63.
32. Dayal, S., A. Mortazavi, K. H. Huynh, R. L. Gerges, and J. J. Shynk. A Cooperative High-Accuracy Localization Algorithm for Improved Road Workers' Safety. in *Asilomar Conference on Signals, Systems and Computers* (2013).
33. Sakhakarmi, S., J. W. Park, and A. Singh. Tactile-based wearable system for improved hazard perception of

worker and equipment collision. *Autom Constr*, 2021. 125.
34. Ibrahim, M., and O. Moselhi. Onsite data acquisition using low cost open source microcontroller. in *ISARC 2013 - 30th International Symposium on Automation and Robotics in Construction and Mining, Held in Conjunction with the 23rd World Mining Congress* (Canadian Institute of Mining, Metallurgy and Petroleum, 2013). doi:10.22260/isarc2013/0044.
35. Vorhes, G., and D. Noyce. Low-Cost Connected Work Zone Devices with Simplified Architecture. in *Transportation Research Record* (SAGE Publications Ltd, 2022). vol. 2676.
36. Nnaji, C., A. Jafarnejad, and J. Gambatese. Effects of Wearable Light Systems on Safety of Highway Construction Workers. *Practice Periodical on Structural Design and Construction*, 2020. 25.
37. Wang, J., and S. N. Razavi. Low False Alarm Rate Model for Unsafe-Proximity Detection in Construction. *Journal of Computing in Civil Engineering*, 2016. 30.
38. Yassin, M., and E. Rachid. A survey of positioning techniques and location based services in wireless networks. *2015 IEEE International Conference on Signal Processing, Informatics, Communication and Energy Systems, SPICES 2015*, 2015. doi:10.1109/SPICES.2015.7091420.
39. Shoval, N. Tracking technologies and urban analysis. *Cities*, 2008. 25.
40. Talmaki, S., S. M. Asce, V. R. Kamat, and M. Asce. Multi-sensor monitoring for real-time 3D visualization of construction equipment. in *ISARC 2013 - 30th International Symposium on Automation and Robotics in Construction and Mining, Held in Conjunction with the 23rd World Mining Congress* (2013).
41. Jiao, R. J., and J. Y. Tsai. Systems Analysis and Design of a Smart Traffic Service System for Predictive and Smarter Mobility and Safety in Roadway Work Zones. in *IEEE International Conference on Industrial Engineering and Engineering Management* (2018).
42. Lu, M., W. Chen, X. Shen, H. C. Lam, and J. Liu. Positioning and tracking construction vehicles in highly dense urban areas and building construction sites. *Autom Constr*, 2007. 16.
43. Pradhananga, N., and J. Teizer. Automatic spatio-temporal analysis of construction site equipment operations using GPS data. *Autom Constr*, 2013. 29.
44. Ometov, A., *et al.* A Survey on Wearable Technology: History, State-of-the-Art and Current Challenges. *Computer Networks* Preprint at https://doi.org/10.1016/j.comnet.2021.108074 (2021) vol. 193.
45. Torrent, D. G., and C. H. Caldas. Methodology for Automating the Identification and Localization of Construction Components on Industrial Projects. *Journal of Computing in Civil Engineering*, 2009. 23.
46. Saeki, M., and M. Hori. Development of an Accurate Positioning System Using Low-Cost L1 GPS Receivers. *Computer-Aided Civil and Infrastructure Engineering*, 2006. 21.
47. Yang, C. C., and Y. L. Hsu. A review of accelerometry-based wearable motion detectors for physical activity monitoring. *Sensors (Basel)*, 2010. 10.
48. Herrera-May, A. L., *et al.* Recent Advances of MEMS Resonators for Lorentz Force Based Magnetic Field Sensors: Design, Applications and Challenges. *Sensors 2016, Vol. 16, Page 1359*, 2016. 16.
49. Sabeti, S., O. Shoghli, and H. Tabkhi. Toward Wi-Fi-Enabled Real-Time Communication for Proactive Safety Systems in Highway Work Zones: A Case Study. in *Construction Research Congress 2022: Computer Applications, Automation, and Data Analytics - Selected Papers from Construction Research Congress 2022* (2022).
50. Nnaji, C., J. Gambatese, and H. W. Lee. Work Zone Intrusion: Technology to Reduce Injuries and Fatalities. *Prof Saf*, 2018. 63.
51. Zhu, X., and D. Fan. Vehicle Intrusion Warning System in Work Zone within Vehicle Infrastructure Integration Environment. in *Vehicle Intrusion Warning System in Work Zone within Vehicle Infrastructure Integration Environment* (2020).
52. Thapa, D., and S. Mishra. Using worker's naturalistic response to determine and analyze work zone crashes in the presence of work zone intrusion alert systems. *Accid Anal Prev*, 2021. 156.
53. Theiss, L. A., T. Lindheimer, and G. L. Ullman. Closed course performance testing of a work zone intrusion alarm system. *Transp Res Rec*, 2018. 2672.
54. Qiao, F., R. Rahman, Q. Li, and L. Yu. Safe and Environment-Friendly Forward Collision Warning Messages in the Advance Warning Area of a Construction Zone. *International Journal of Intelligent Transportation Systems Research*, 2017. 15.
55. Jumari, M. Z. I., *et al.* Signal Warning Detector (SWAD) for Sustainable Working Environment at Highways. *International Journal of Integrated Engineering*, 2022. 14.
56. Yanli, M., G. Gaofeng, J. Yuan, and Z. Xuesheng. Speed control system analysis of freeway work zone based on ITS. in *Proceedings - 2014 5th International Conference on Intelligent Systems Design and Engineering Applications, ISDEA 2014* (Institute of Electrical and Electronics Engineers Inc., 2014).

doi:10.1109/ISDEA.2014.267.
57. Cui, H., *et al.* Vision-Based Work Zone Safety Alert System in a Connected Vehicle Environment. *Transp Res Rec*, 2023. 2677.
58. Ullman, G. L., M. D. Finley, and L. A. Theiss. Categorization of work zone intrusion crashes. *Transp Res Rec*, 2011. doi:10.3141/2258-07.
59. Li, Y., F. Cheng, and Y. Bai. *Characteristics of Truck-Related Crashes in Highway Work Zones*.
60. Goh, T. H., H. S. Alzraiee, and M. Asce. *Digital Safety Planning and Monitoring in Highway Construction Projects*.
61. Tymvios, N., and S. Oosthuysen. *Distracted Driver Speeds around Work Zones*.
62. Huang, Y., and Y. Bai. Effectiveness of graphic-aided portable changeable message signs in reducing vehicle speeds in highway work zones. *Transp Res Part C Emerg Technol*, 2014. 48.
63. Shi, W., and R. R. Rajkumar. Work Zone Detection for Autonomous Vehicles. in *IEEE Conference on Intelligent Transportation Systems, Proceedings, ITSC* (Institute of Electrical and Electronics Engineers Inc., 2021). vols 2021-September.
64. Rahman, M. M., L. Strawderman, T. Garrison, D. Eakin, and C. C. Williams. Work zone sign design for increased driver compliance and worker safety. *Accid Anal Prev*, 2017. 106.
65. Nnaji, C., A. A. Karakhan, J. Gambatese, and H. W. Lee. Case Study to Evaluate Work-Zone Safety Technologies in Highway Construction. *Practice Periodical on Structural Design and Construction*, 2020. 25.
66. Lin, P.-S., Z. Wang, ; Rakesh Rangaswamy, R. Durga, and T. N. Kolla. Innovative Approaches to Significantly Improve Arterial Work Zone Safety. in *International Conference on Transportation and Development 2023: Transportation Safety and Emerging Technologies - Selected Papers from the International Conference on Transportation and Development 2023* (2023).
67. Chai, H., *et al.* Discussion of active safety countermeasures of motorway work zone. in *IOP Conference Series: Earth and Environmental Science* (Institute of Physics Publishing, 2019). vol. 371.
68. Martin, J., A. Rozas, and A. Araujo. A WSN-Based Intrusion Alarm System to Improve Safety in Road Work Zones. *J Sens*, 2016. 2016.
69. Phanomchoeng, G., R. Rajamani, and J. Hourdos. Directional sound for long-distance auditory warnings from a highway construction work zone. *IEEE Trans Veh Technol*, 2010. 59.
70. Wang, S., A. Sharma, and S. Knickerbocker. Analyzing and improving the performance of dynamic message sign reporting work zone–related congestion. *Transp Res Rec*, 2017. 2617.
71. Zhang, F., and J. A. Gambatese. Highway Construction Work-Zone Safety: Effectiveness of Traffic-Control Devices. *Practice Periodical on Structural Design and Construction*, 2017. 22.
72. Gambatese, J., and F. Zhang. Impact of advisory signs on vehicle speeds in highway nighttime paving project work zones. *Transp Res Rec*, 2016. 2555.
73. Bai, Y., and Y. Li. Determining the drivers' acceptance of EFTCD in highway work zones. *Accid Anal Prev*, 2011. 43.
74. Bai, Y., K. Finger, and Y. Li. Analyzing motorists' responses to temporary signage in highway work zones. *Saf Sci*, 2010. 48.
75. Lopez-Flores, D., *et al.* Implementation of two robotic flagmen controlled by CAN messages to increase the safety of human workers in road maintenance. *Journal of Applied Research and Technology*, 2020. 18.
76. Ščerba, M., T. Apeltauer, and J. Apeltauer. Portable telematic system as an effective traffic flow management in workzones. *Transport and Telecommunication*, 2015. 16.
77. Wavetronix - SmartSensor HD. https://www.wavetronix.com/products/smartsensor-hd.
78. Fan, W., S. Choe, and F. Leite. *Prevention of Backover Fatalities in Highway Work Zones: A Synthesis of Current Practices and Recommendations*. *International Journal of Transportation Science and Technology* · (2014) vol. 3.
79. Brown, H., C. Sun, and T. Cope. Evaluation of mobile work zone alarm systems. *Transp Res Rec*, 2015. 2485.
80. Sakhakarmi, S., and J. W. Park. Improved intrusion accident management using haptic signals in roadway work zone. *J Safety Res*, 2022. 80.
81. Wang, J., *et al.* Crash prediction for freeway work zones in real time: A comparison between Convolutional Neural Network and Binary Logistic Regression model. *International Journal of Transportation Science and Technology*, 2022. 11.
82. Dehman, A., and B. Farooq. Are work zones and connected automated vehicles ready for a harmonious coexistence? A scoping review and research agenda. *Transp Res Part C Emerg Technol*, 2021. 133.
83. Darwesh, A., D. Wu, M. Le, and S. Saripalli. Building a Smart Work Zone Using Roadside LiDAR. in *IEEE Conference on Intelligent Transportation Systems, Proceedings, ITSC* (2021).

84. Rao, A. S., *et al.* Real-time monitoring of construction sites: Sensors, methods, and applications. *Autom Constr*, 2022. 136.
85. Pučko, Z., N. Šuman, and D. Rebolj. Automated continuous construction progress monitoring using multiple workplace real time 3D scans. *Advanced Engineering Informatics*, 2018. 38.
86. Adarsh, S., S. M. Kaleemuddin, D. Bose, and K. I. Ramachandran. Performance comparison of Infrared and Ultrasonic sensors for obstacles of different materials in vehicle/ robot navigation applications. in *IOP Conference Series: Materials Science and Engineering* (Institute of Physics Publishing, 2016). vol. 149.
87. Choe, S., F. Leite, D. Seedah, R. Associate, and C. Caldas. Evaluation of sensing technology for the prevention of backover accidents in construction work zones. *Journal of Information Technology in Construction*, 2014. 19.
88. Paidi, V., H. Fleyeh, J. Håkansson, and R. G. Nyberg. Smart parking sensors, technologies and applications for open parking lots: A review. *IET Intelligent Transport Systems* Preprint at https://doi.org/10.1049/iet-its.2017.0406 (2018) vol. 12.
89. Odat, E., J. S. Shamma, and C. Claudel. Vehicle Classification and Speed Estimation Using Combined Passive Infrared/Ultrasonic Sensors. *IEEE Transactions on Intelligent Transportation Systems*, 2018. 19.
90. Park, J. W., X. Yang, Y. K. Cho, and J. Seo. Improving dynamic proximity sensing and processing for smart work-zone safety. *Autom Constr*, 2017. 84.
91. Qiao, Y., and F. Qiao. Person-to-Infrastructure (P2I) Wireless Communications for Work Zone Safety Enhancement. in *2012 15th International IEEE Conference on Intelligent Transportation Systems* (2012).
92. Park, J., ; E Marks, ; Y K Cho, and W. Suryanto. Mobile Proximity Sensing Technologies for Personnel and Equipment Safety in Work Zones. in *Congress on Computing in Civil Engineering, Proceedings* (2015).
93. Park, J., Y. K. Cho, and S. K. Timalsina. Direction Aware Bluetooth Low Energy Based Proximity Detection System for Construction Work Zone Safety. in *ISARC 2016 - 33rd International Symposium on Automation and Robotics in Construction* (2016).
94. Kim, K., I. Jeong, and Y. K. Cho. Signal Processing and Alert Logic Evaluation for IoT–Based Work Zone Proximity Safety System. *J Constr Eng Manag*, 2023. 149.
95. Teizer, J. Wearable, wireless identification sensing platform: Self-Monitoring Alert and Reporting Technology for Hazard Avoidance and Training (SmartHat). *Journal of Information Technology in Construction (ITcon)*, 2015. 20.
96. Malinovskiy, Y., N. Saunier, and Y. Wang. Analysis of pedestrian travel with static bluetooth sensors. *Transp Res Rec*, 2012. doi:10.3141/2299-15.
97. Park, J., E. Marks, Y. K. Cho, and W. Suryanto. Performance Test of Wireless Technologies for Personnel and Equipment Proximity Sensing in Work Zones. *J Constr Eng Manag*, 2016. 142.
98. Yang, X., J. Park, and Y. K. Cho. Adaptive Signal-Processing for BLE Sensors for Dynamic Construction Proximity Safety Applications. in *Construction Research Congress 2018: Safety and Disaster Management - Selected Papers from the Construction Research Congress 2018* (2018).
99. Izzatdin, A., J. Jaafar, N. Natrah, and B. Ismail. Roadside worker detection and alert system using RFID. *ARPN Journal of Engineering and Applied Sciences*, 2015.
100. Mezzanotte, P., V. Palazzi, F. Alimenti, and L. Roselli. Innovative RFID Sensors for Internet of Things Applications. *IEEE Journal of Microwaves*, 2021. 1.
101. Hallowell, M. R., J. Teizer, and W. Blaney. Application of Sensing Technology to Safety Management. in *Construction Research Congress 2010: Innovation for Reshaping Construction Practice - Proceedings of the 2010 Construction Research Congress* (2010).
102. Costa, F., *et al.* A review of rfid sensors, the new frontier of internet of things. *Sensors* Preprint at https://doi.org/10.3390/s21093138 (2021) vol. 21.
103. Landaluce, H., *et al.* A review of iot sensing applications and challenges using RFID and wireless sensor networks. *Sensors (Switzerland)* Preprint at https://doi.org/10.3390/s20092495 (2020) vol. 20.
104. Kim, K., H. Kim, and H. Kim. Image-based construction hazard avoidance system using augmented reality in wearable device. *Autom Constr*, 2017. 83.
105. Han, W., E. White, ; Mike Mollenhauer, and N. Roofigari-Esfahan. A Connected Work Zone Hazard Detection System for Roadway Construction Workers. in *Computing in Civil Engineering* (2019).
106. *Minnesota Temporary Traffic Control Field Manual*. (2018).
107. Venthuruthiyil, S. P., D. Thapa, and S. Mishra. Towards smart work zones: Creating safe and efficient work zones in the technology era. *J Safety Res*, 2023. doi:10.1016/j.jsr.2023.08.006.
108. Ozan, E. Requirements analysis for the system level design of smart work zones. in *14th Annual Conference System of Systems Engineering, SoSE* (2019).

109. Yang, X., and N. Roofigari-Esfahan. Vibrotactile Alerting to Prevent Accidents in Highway Construction Work Zones: An Exploratory Study. *Sensors*, 2023. 23.
110. Teizer, J., B. S. Allread, C. E. Fullerton, and J. Hinze. Autonomous pro-active real-time construction worker and equipment operator proximity safety alert system. *Autom Constr*, 2010. 19.
111. Tosi, J., F. Taffoni, M. Santacatterina, R. Sannino, and D. Formica. Performance evaluation of bluetooth low energy: A systematic review. *Sensors (Switzerland)* Preprint at https://doi.org/10.3390/s17122898 (2017) vol. 17.
112. Rosenberger, P., *et al.* Sequential lidar sensor system simulation: a modular approach for simulation-based safety validation of automated driving. *Automotive and Engine Technology 2020 5:3*, 2020. 5.
113. Kang, J., *et al.* Modular and Reconfigurable Stretchable Electronic Systems. *Adv Mater Technol*, 2019. 4.
114. Siekkinen, M., M. Hiienkari, J. Nurminen, and J. Nieminen. How Low Energy is Bluetooth Low Energy? Comparative Measurements with ZigBee/802.15.4. in *WCNC 2012 Workshop on Internet of Things Enabling Technologies, Embracing Machine-to-Machine Communications and Beyond* (IEEE, 2012).
115. Nnaji, C., H. W. Lee, A. Karakhan, and J. Gambatese. Developing a Decision-Making Framework to Select Safety Technologies for Highway Construction. *J Constr Eng Manag*, 2018. 144.
116. Sabeti, S., O. Shoghli, N. Morris, and H. Tabkhi. Wearable Technology for Highway Maintenance and Operation Safety: A Survey of Workers' Perception and Preferences. in *Proceedings of the International Symposium on Automation and Robotics in Construction* (International Association for Automation and Robotics in Construction (IAARC), 2022). vols 2022-July.
117. Burg, A., A. Chattopadhyay, and K. Y. Lam. Wireless communication and security issues for cyber- physical systems and the internet-of-things. *Proceedings of the IEEE*, 2018. 106.